# Training for Technology:

# Adoption and Productive Use of Generative AI in Legal Analysis

Benjamin M. Chen[1] Hong Bao[2]

## Abstract

*Can targeted user training unlock the productive potential of generative artificial intelligence in professional settings? We study this question using a randomized experiment in which 164 law students completed an issue-spotting examination under one of three conditions: no GenAI access, optional access to a large language model (LLM), or LLM access with a brief training intervention.*

*Untrained LLM access proved counterproductive: relative to participants without any LLM access, untrained users wrote significantly shorter answers, committed more case misstatements, and scored marginally lower, though most differences fall short of conventional significance. Training reversed this pattern. Trained participants adopted the LLM at higher rates (41% vs. 26%; p = 0.044), scored 0.27 grade points higher than*

[1] Associate Professor and Gallant Ho Outstanding Young Professor, University of Hong Kong Faculty of Law. Comments and criticisms are welcome and should be directed to benched@hku.hk.

[2] PhD student, University of Hong Kong Faculty of Law.

*untrained users — roughly one fine grade — ($p = 0.027$), and stated applicable rules more accurately ($p = 0.014$).*

*Principal stratification analysis suggests training operates primarily through adoption rather than effectiveness — the adoption lower bound (1.06) exceeds the effectiveness upper bound (0.42) at strict mean dominance — though confidence intervals are wide.*

*More broadly, these findings challenge the view that GenAI primarily benefits lower-skilled workers: without training, higher-ability practitioners opt out while lower-ability users adopt but unproductively. Realizing GenAI's productivity gains requires investment in both access and instruction.*

## Introduction

The advent of generative artificial intelligence (GenAI) represents one of the most significant technological transformations for the learned professions since the computerization of white-collar labor in the late twentieth century. Specifically, large language models (LLMs) exhibit capabilities that range from drafting technical documents to dispensing high-stakes advice. Yet the deployment of GenAI has proceeded unevenly across and within disciplines. This gap between rhetoric and reality reflects a fundamental tension: while GenAI promises to raise labor productivity and complement human expertise, its eventual impact depends on where, how, and by whom the technology is deployed.

This article asks whether user training can enhance human productivity by increasing use of GenAI—or the effectiveness of such use—for applications where the technology may not be fully reliable and where mistakes can be especially costly. We present evidence suggesting that it can.

Our experiment took the form of a mock examination administered to 164 law undergraduate (LLB) and graduate (JD) students enrolled in a core module in contract law at the University of Hong Kong. Participants were randomly assigned to one of three groups. The first group did not have access to any GenAI assistance. The second group was permitted to use an LLM but did not receive any guidance on how they should use it. The third group was not only permitted to use an LLM but also watched an instructional video and took a short quiz on how GenAI can help identify and analyze legal issues raised by a fact pattern.

Overall, access to an LLM did not noticeably improve performance on the examination absent user training. Indeed, if anything, participants who had access to an LLM without any intervention wrote less and scored worse than those who did not have such access although the latter decline is small and statistically insignificant. Compared to participants who only had access to an LLM, participants who received the intervention reported substantially higher rates of GenAI usage and achieved, on average, scores that were better by nearly one fine grade.

While many studies have confirmed the ability of GenAI tools to execute cognitively burdensome tasks, unsupervised professional practice by GenAI remains a distant possibility. More attention should therefore be paid to how humans call on, relate to, and collaborate with GenAI. Our findings indicate that targeted instruction on how GenAI can and ought to be employed for certain kinds of applications may be critical in fully realizing the benefits of GenAI for knowledge-intensive fields.

## 1. Background

### 1.1. GenAI and its Impacts on the Labor Market

Economists have long recognized that major technological revolutions affect workers unevenly across the labor market. In the early 2000s, efforts to understand the consequences of computerization led to skill-based and task-based frameworks that distinguished between routine and non-routine work. These literatures argued that computerization primarily substituted for low or middle-skill workers performing routine tasks, while complementing higher-skill workers engaged in cognitive, non-routine activities (Acemoglu and Restrepo 2020; Acemoglu and Autor 2011; Autor et al. 2003).

The emergence of artificial intelligence complicates this framework. Unlike prior computing systems—characterized by “faultless and nearly costless execution of routine, procedural tasks”—AI systems exhibit capabilities associated with acquiring tacit knowledge and exercising forms of expert judgment (Autor 2024). Empirical evidence suggests that large language models (LLMs) may therefore expose higher-income and higher-skill occupations to technological change to a greater extent than previous innovations (Eloundou et al. 2023). This raises the possibility that AI differs fundamentally from earlier technologies by displacing rather than complementing human expertise.

Much of the recent empirical literature, however, paints a more limited picture. Acemoglu (2025) asserts that observed productivity gains from AI are concentrated in relatively “easy tasks” across domains such as customer support, software development, and professional writing. Related studies likewise document levelling effects, whereby AI disproportionately benefits less-skilled workers and compresses performance differences rather than amplifying skill premia (Brynjolfsson et al. 2023; Dell′Acqua et al. 2023; Noy and Zhang 2023; Peng et al. 2023; Cui et al. 2025). These findings suggest continuity rather than rupture with earlier patterns of task-based technological change.

Whether this pattern holds for boundary applications — tasks that are judgment-intensive, error-sensitive, and context-dependent — is less clear. In such settings, the effects of generative AI depend critically on the nature of the task and on how users interact with the technology (Dellʹ'Acqua et al. 2023).

### 1.2. GenAI and the Legal Profession

Legal work exemplifies such boundary applications. A growing body of research demonstrates that GenAI can perform a wide range of legal tasks, including drafting contractual clauses (Lam et al. 2023), summarizing judicial opinions (Bauer et al. 2023), and even producing case commentary (Engel and Kruse 2025). Controlled studies corroborate the productivity promise. Schwarcz et al. (2025) find that the use of retrieval-augmented generation (RAG) and AI reasoning models improves the time-adjusted quality of law students' work by more than 50% across tasks ranging from drafting legal memoranda to analyzing complaints. Nielsen et al. (2024) report that AI-generated highlighting of legal filings reduces the time required for law students to understand and answer questions about complaints by up to 30%. Choi et al. (2024) estimate time savings of 10% to 30% when law students use GPT-4 to perform basic legal tasks such as drafting contracts or memoranda, with larger gains accruing to lower-performing students.

Practice tells a different story. A 2025 survey of more than 2,800 lawyers reports that only 21% of law firms currently use GenAI tools (American Bar Association 2025). Another survey shows despite broad access to AI tools, less than a third of the law professionals report confidence in using the tools (Denniston and Duffy 2026). Practitioners appear particularly cautious in contexts where accuracy is critical. Adoption rates for legal drafting lag behind those for document review and legal research (LexisNexis 2024; Thomson Reuters 2025). Notably, in one survey, 74% of firms identify concerns about hallucinated or fabricated content as a major barrier to deployment (LexisNexis 2024). This concern is well-founded. Dahl et al. (2024) put hallucination rates at as high as 58% for tasks such as identifying the legal rule established by a judicial decision, and at least 851 court opinions have addressed the submission of AI-fabricated content by lawyers or litigants (Charlotin, n.d.).

The gap between what GenAI can do and how lawyers use it motivates this study. Where doctrinal ambiguity is high and errors are costly, the expected value of GenAI use depends critically on how users interact with the technology. Effective human-AI collaboration, in such settings, cannot be assumed — it may need to be taught.

### 1.3. User Training as a Mechanism for GenAI Adoption and Productivity

At the current stage of technological development, GenAI primarily augments rather than automates professional labor: it assists human decision-makers rather than replacing their judgment (Agrawal et al. 2026). This study examines whether the legal profession can realize greater productivity gains from GenAI through appropriate user interventions, particularly for high-skill workers engaged in non-routine tasks.

We begin by sketching a simple conceptual framework in which GenAI can enhance human productivity through two distinct channels: by expanding the scope of GenAI use—that is, the set of tasks for which users are willing to adopt the technology—and by improving the effectiveness of use conditional on adoption. We hypothesize that user training increases the likelihood that high-skill professionals will employ GenAI for complex, non-routine tasks and also improves how users interact with the technology once adoption occurs.

Survey evidence consistently identifies the lack of training as a primary barrier to GenAI adoption (Humlum and Vestergaard 2024; Freitas 2025), but not much research has been done to isolate the marginal contribution of training to productivity. The most relevant evidence from the legal profession comes from a pilot field study by Chien and Kim (2025), in which legal aid lawyers were given temporary access to paid GenAI tools, with a subset also receiving concierge support that included peer use cases and office hours. Participants assigned to this concierge support condition reported higher perceived productivity and stronger intentions to adopt GenAI in the future. Outcomes, however, were self-reported rather than objectively measured. Related experimental studies involving law students featured instruction in using GenAI but do not isolate the marginal contribution of user training (Choi et al. 2024; Choi and Schwarcz 2025; Schwarcz et al. 2025). For example, Choi and Schwarcz (2025) compare students without access to GenAI to students who use GenAI following extensive preparation, finding performance gains on multiple-choice questions but not on essay-based tasks.

Our study differs from prior research in several important respects. First, we recognize that lawyers today typically have a choice about whether to use GenAI. Accordingly, access to an LLM in our experiment was optional rather than required. Second, we focus on tasks that demand both professional judgment and accuracy by situating our experiment in the context of an issue-spotting examination that requires participants to identify and analyze legal issues raised by a complex fact pattern. Third, we explicitly examine the role of user training in shaping both adoption decisions and the quality of professional output. To this end, our experimental design includes two treatment groups with identical access to an LLM, differing only in whether participants received a brief instructional intervention in the form of a training video and quiz. Together, these design choices allow us to more clearly assess how user training mediates the productivity effects of GenAI and to draw broader implications for the future organization of the learned professions.

## 2. Theoretical Motivation

This section develops a stylized model to clarify how training can enhance productivity by enhancing the scope or the effectiveness of GenAI use. Consider a lawyer with ability $\theta$ performing a legal task indexed by its complexity $c \in [0,1]$. The lawyer can choose whether to employ GenAI. Individual productivity is

$$Y_T = \theta + D \cdot (1 - \theta)[A \cdot e_T - L(c, e_T)]$$

where $\theta \in [0,1]$ represents professional ability, $D \in \{0,1\}$ represents the adoption decision, $A$ represents the base capability of the technology, $T \in \{0,1\}$ represents exposure to training, $e_T \in [0,1]$ represents effectiveness of use as a function of training, and

$$L(c, e_T) = p(c, e_T) \cdot l(c)$$

is the expected loss from errors caused by using GenAI, where $p(c, e_T)$ is the probability of error and $l(c)$ is the cost of error on task $c$. Error rates are higher on more complex tasks, i.e. $\frac{\partial p}{\partial c} > 0$, and more effective use reduces error rates, i.e. $\frac{\partial p}{\partial e_T} < 0$. Errors are also costlier for more complex tasks, i.e. $\frac{\partial l}{\partial c} > 0$.

The net productivity gain from using GenAI is thus:

$$\Delta Y_T(c, e_T, \theta) = (1-\theta) \cdot [A \cdot e_T - p(c, e_T) \cdot l(c)].$$

All else equal, a higher ability lawyer benefits less from GenAI adoption than a lower ability lawyer. Moreover, for boundary applications where $c$ is large, the loss term can dominate making $\Delta Y_T < 0$ even when $A \cdot e_T$ is substantial. Effectiveness here is independent of ability $\theta$. The marginal value of effectiveness, however, does depend on ability. In general, higher effectiveness increases productivity conditional on adoption,

$$\frac{\partial \Delta Y}{\partial e_T} = (1-\theta)\left[A - \frac{\partial p}{\partial e_T} \cdot l(c)\right] > 0$$

but it does so more for lower ability compared to higher ability individuals,

$$\frac{\partial^2 \Delta Y}{\partial e_T \partial \theta} = -\left[A - \frac{\partial p}{\partial e_T} \cdot l(c)\right] < 0.$$

Under this model, a lawyer of ability $\theta$ uses GenAI for task $c$ if

$$(1-\theta) \cdot [A \cdot e_T - p(c, e_T) \cdot l(c)] > k_T \tag{1}$$

where $k_T > 0$, the cost of adoption, is a function of training. $k_T$ here can include time and money, risk premium, and psychological barriers. We assume that training does not hurt effectiveness, i.e., $e_1 \geq e_0$ and does not increase the cost of adoption, i.e. $k_1 \leq k_0$. Equation (1) then gives rise to a natural distinction between always users—those who would adopt GenAI for a given task even in the absence of training—and induced users, whose adoption is triggered by training. Assuming $A \cdot e_0 - p(c, e_0) \cdot l(c) > 0$, always users are those for whom

$$\theta < 1 - \frac{k_0}{A \cdot e_0 - p(c, e_0) \cdot l(c)}$$

and induced users are those for whom

$$1 - \frac{k_0}{A \cdot e_0 - p(c, e_0) \cdot l(c)} \leq \theta < 1 - \frac{k_1}{A \cdot e_1 - p(c, e_1) \cdot l(c)}.$$

For always users, training operates primarily through improvements in effectiveness, reducing error rates and increasing net productivity conditional on adoption. For induced users, training operates through the extensive margin by lowering adoption costs or mitigating perceived risks, thereby expanding the set of tasks for which GenAI use becomes attractive.

## 3. Experimental Design

To explore the effect of user training on the scope and effectiveness of GenAI use, we ran an experiment featuring a single-factor, between-subjects design.

Participants were invited to take a mock examination on contract law. The examination was open-book, and all participants were given one hour and fifteen minutes to submit their answers, with fifteen minutes being reserved for reading and sixty minutes for typing. The examination question required participants to identify the legal issues presented by a hypothetical fact pattern and to analyze them to reach a conclusion about the merits of a party's argument. This type of examination is commonly known as an issue-spotter and tests, among other things, candidates' knowledge of black-letter law, their sensitivity to normatively relevant facts, their capacity to navigate legal ambiguity and their judgment in assessing the strength of competing arguments. The examination here was adapted from an official, supplementary examination administered in 2021.[3] The marking rubric enumerates four major issues, namely, formation of the contract, misrepresentation, promissory estoppel and breach of implied terms. A total of twelve sub-issues were listed under these four headings.

Participants were randomly assigned to one of three groups for the mock examination. Those in the first group, i.e. Group 1, could only query a traditional legal database, Westlaw, through the internet and were not permitted to use any GenAI tools. Participants in the second and third groups, i.e. Groups 2 and 3, could query both Westlaw and DeepSeek. Participants in Group 2 did not receive any guidance on using an LLM for issue-spotting. By contrast, participants in Group 3 watched a nine-minute, thirty-second video entitled "Using Large Language Models to Help Your Legal Analysis (Figure 1)."[4] They also had to complete a five-question multiple-choice quiz about the contents of the video before starting the examination.

**Figure 1:** A Screenshot of the Training Video

[3] The supplementary examination was taken by a small number of students, none of whom were eligible for this study, and has not been made available as part of the University of Hong Kong Law Library's collection of past year papers.

[4] This material was created using Youyan 3D, a video generation platform.

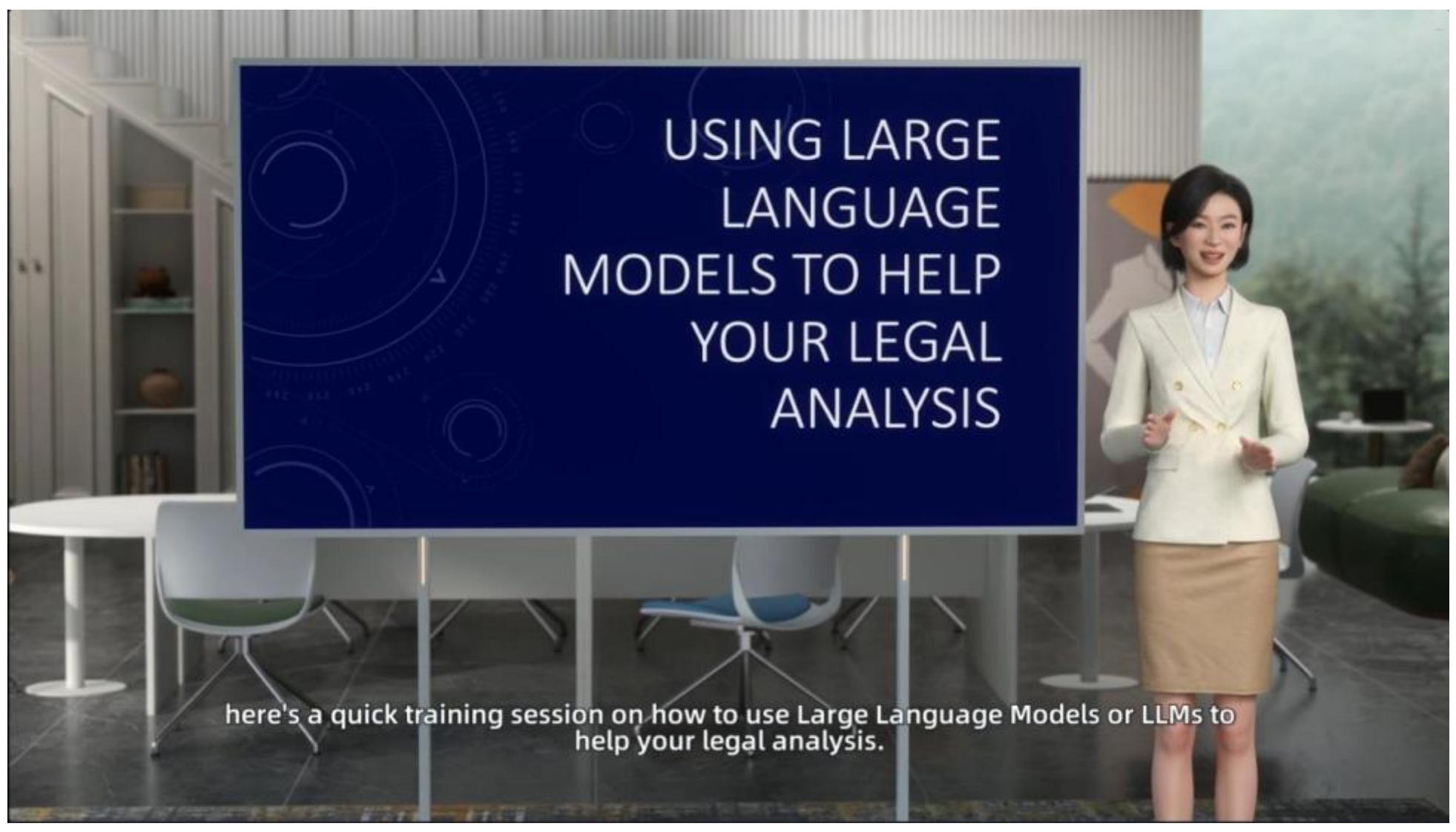

**Alt Text**: Screenshot of the training video "Using Large Language Models to Help Your Legal Analysis." A virtual avatar presents instructional content on prompt engineering and verification strategies for legal analysis.

Inspired by the duty of competence requirement, which includes informed decision and verification (Conley 2025), the training video at the core of our intervention consisted of two parts. The first part described how LLMs could be used to help analyze a legal problem. Recommendations, drawn from prior research, included repeating the prompt several times so that the model generates more insights, breaking the task down into intermediate steps to facilitate chain-of-thought reasoning, and giving the model both positive and negative feedback so that it returns more tailored completions (Schwarcz and Choi 2023). This part of the video also cautioned that: "Generic LLMs produce responses based on probabilistic models of language and there is no guarantee that the text they generate make sound or accurate claims." It advised that an LLM should be employed to assist on a legal task only if the user has sufficient knowledge to check the accuracy of output, that the user should ask the LLM to give ideas or directions for solving a legal problem rather than relying on it for a solution, and that the user should always double-check output from an LLM, including all facts, rules, and cases referred to. The second part of the video then showed how these techniques and strategies could be applied in practice. The demonstration made use of a stylized fact pattern implicating a legal issue irrelevant to the ones tested in the examination. By visualizing the interaction between a human and the LLM, the video demonstrated the steps participants could take if they chose to take advantage of GenAI for the examination.

After taking the mock examination, all participants were surveyed about their experiences with and attitudes toward GenAI tools. Specifically, they were asked whether they had received any previous training in using LLMs for legal research or legal analysis, whether they used DeepSeek in the examination, whether they agreed that having access to LLMs would help them in law examinations, and whether they agreed that students should be allowed to have access to LLMs for law examinations.

Several directional and non-directional hypotheses were enumerated in a pre-registration plan lodged by the authors on the Open Science Framework website.[5]

## 4. Data Collection

Participants were recruited from students enrolled in undergraduate, i.e. LL.B., and graduate, i.e. J.D., contract law courses at the University of Hong Kong. The mock examination was scheduled for 30 November 2025.[6] An email inviting students to participate was sent approximately two months before their official mid-year examination on 15 December 2025. Students were promised HKD 100 Starbucks gift cards and feedback on the mock examination if they completed the study.

[5] Hypotheses tests reported in the article that were not pre-registered on the Open Science Framework will be declared as such. Hypotheses tests that were pre-registered but omitted from the article for reasons of brevity are reported in the appendix.

A total of 213 students registered for the mock examination. Among them, 171 were LL.B. students and 42 were J.D. students. 60 participants were randomly assigned to Group 1, 76 to Group 2, and 77 to Group 3. All participants were instructed to bring a laptop computer to the examination. Participants were given different reporting locations depending on their group numbers. To preserve the stable unit treatment value assumption and to minimize differential attrition across groups, students were not informed in advance that some of them would be permitted to access DeepSeek during the examination.

Group 1 was split across two small classrooms, each supervised by one invigilator. Group 2 was in one large classroom supervised by two invigilators. Group 3 was in another large classroom supervised by two invigilators.[7] Invigilators implemented the protocol to be followed for the classrooms they were supervising. Recall that Group 1 could only access Westlaw. Group 2 could access Westlaw and DeepSeek but did not receive any training on how an LLM might be helpful for legal analysis. Group 3 could access Westlaw and DeepSeek and, in addition, received the training intervention. For Groups 2 and 3, the option of consulting DeepSeek was stated on the front page of the question paper and reiterated by invigilators before the examination commenced. The same information was also displayed on electronic screens at the front of the classrooms (**Figure 2**).

---

[6] Around two weeks before the mock examination, the most important English language newspaper in Hong Kong, the South China Morning Post, ran stories about the retraction of a journal article authored by a prominent professor at the University of Hong Kong and his PhD student. 20 of 61 references in the article did not exist and were hallucinations by GenAI.

[7] Invigilators were selected to be demographically similar to one another to minimize any possibility of stereotype threat.

**Figure 2:** Information Displayed on Electronic Screens in Classrooms

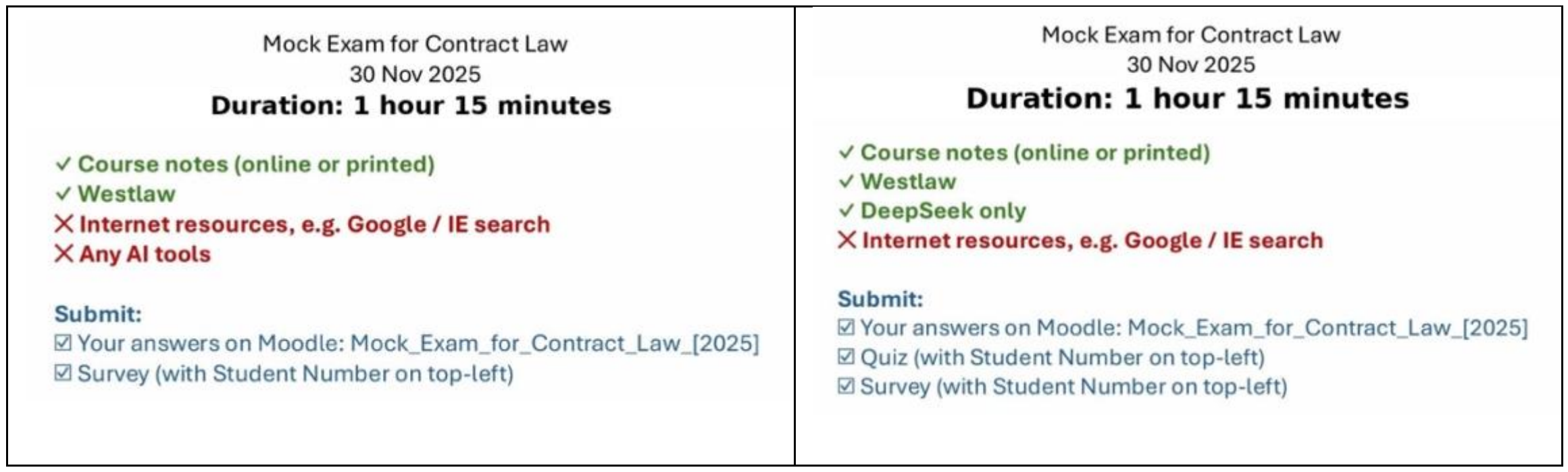

**Alt Text:** Resource access notices displayed on electronic screens in examination classrooms. The left panel shows instructions for Group 1, permitting access to course notes and Westlaw only. The right panel shows instructions for Groups 2 and 3, additionally permitting access to DeepSeek.

At the end of the mock examination, answers were submitted electronically through Moodle, a course management platform. Participants were then asked to fill out the post-examination questionnaire. Participants were identified solely by their university student numbers. Ultimately, 164 of the 213 participants who signed up finished the study: 49 from Group 1, 57 from Group 2, and 58 from Group 3.[8] All participants used the maximum amount of time allotted for the examination.

# 5. Results

## 5.1. Comparing DeepSeek Use

The main outcomes variables for the study are self-reported use of DeepSeek and performance on the mock examination. Self-reported use is retrieved directly from the post-examination questionnaire and is recorded as a binary variable. We hypothesized that the training intervention would increase the rate of self-reported use in Group 3 compared to Group 2.

**Figure 3** plots self-reported use by group. Of the 57 participants in Group 2, 15 declared that they used DeepSeek during the examination, giving an adoption rate of 26.32%. By comparison, 24 of 58 participants in Group 3 declared such use, giving an adoption rate of 41.38%. The difference of 15.06% points is statistically significant ($p$ = 0.044*, one-tailed $z$-test).[9]

[8] One participant originally assigned to Group 3 was ten minutes late and missed the training video. This participant was allowed to finish the examination in Group 3's classroom but is labelled as belonging to Group 2.

[9] As the pre-registration prescribed a $t$-test for all group-wise comparisons and neglected to specify a $z$-test for comparing proportions, the $p$-value for a one-tailed $t$-test is also reported here as 0.045*.

**Figure 3**: Rates of Self-Reported Use of DeepSeek

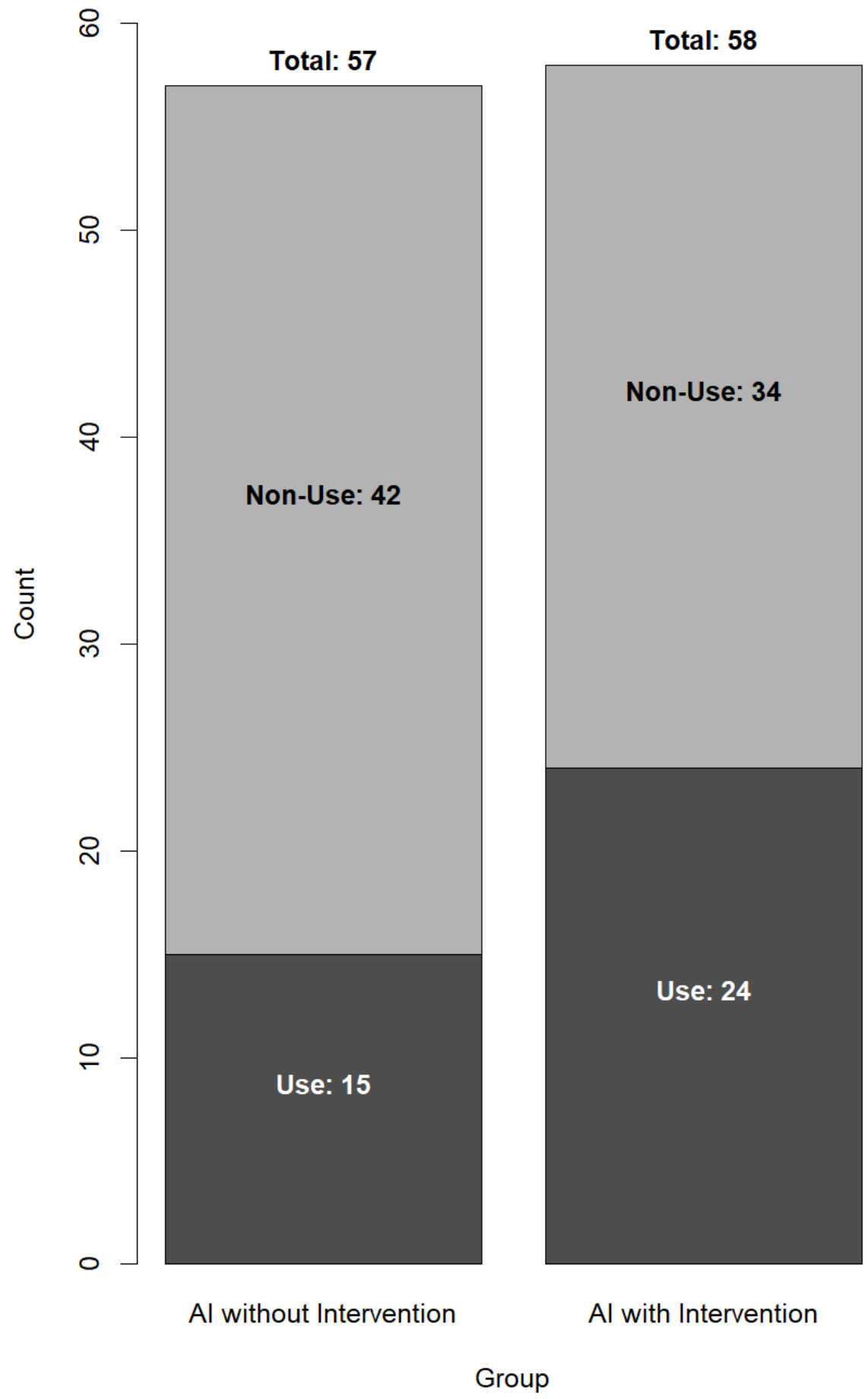

**Alt Text:** Stacked bar chart comparing two groups: "AI without Intervention" (total N=57) and "AI with Intervention" (total N=58). Each bar is split into "Use" (dark gray) and "Non-Use" (light gray) counts. For "AI without Intervention": Use=15, Non-Use=42. For "AI with Intervention": Use=24, Non-Use=34. The y-axis is labeled "Count" and ranges from 0 to 60.

### 5.2. Comparing Examination Performance

Performance is scored according to a marking rubric. All 164 validly submitted answers were anonymized and shuffled before being graded. For each answer, the grader marked whether an issue had been spotted or omitted. If an issue was omitted, the score for that issue was recorded as "NA." If the issue was spotted, the score for that issue may range from zero to two, three or five, depending on the maximum number of sub-issues available. To be clear, then, an "NA" is distinguished from a zero by the fact that in the former case the answer did not address the issue at all whereas in the latter case the issue was considered but in a way that was wholly off the mark. The total score for an answer is the sum of scores across all four issues tested. For the purposes of the total score, an NA counts as zero. Two measures of qualitative performance can be derived. The first is a grade point that corresponds to the letter grade achieved based on the total score. Grade point is a discrete variable that takes on values ranging from 1 to 4.3. The second is the number of issues missed. Number of issues missed is a discrete variable that assumes integer values between 0 and 4.

The grade point distribution for all participants is presented in **Table 1**. **Figure 4** overlays the distribution density of grade point by groups. Graphically, there appears to be a right-shift in the distribution between Groups 1 and 2 and Group 3.

**Table 1:** Distribution of Letter Grades and Grade Points

| Letter Grade | Grade Point | Count |
|---|---|---|
| D | 1 | 9 |
| D+ | 1.3 | 11 |
| C- | 1.7 | 25 |
| C | 2 | 25 |
| C+ | 2.3 | 28 |
| B- | 2.7 | 21 |
| B | 3 | 19 |
| B+ | 3.3 | 13 |
| A- | 3.7 | 9 |
| A | 4 | 2 |
| A+ | 4.3 | 2 |

**Alt Text:** Distribution of letter grades and corresponding grade points among 164 students. Counts are highest for C+ (28), C (25), and C- (25).

**Figure 4:** Distribution Density of Grade Point by Groups

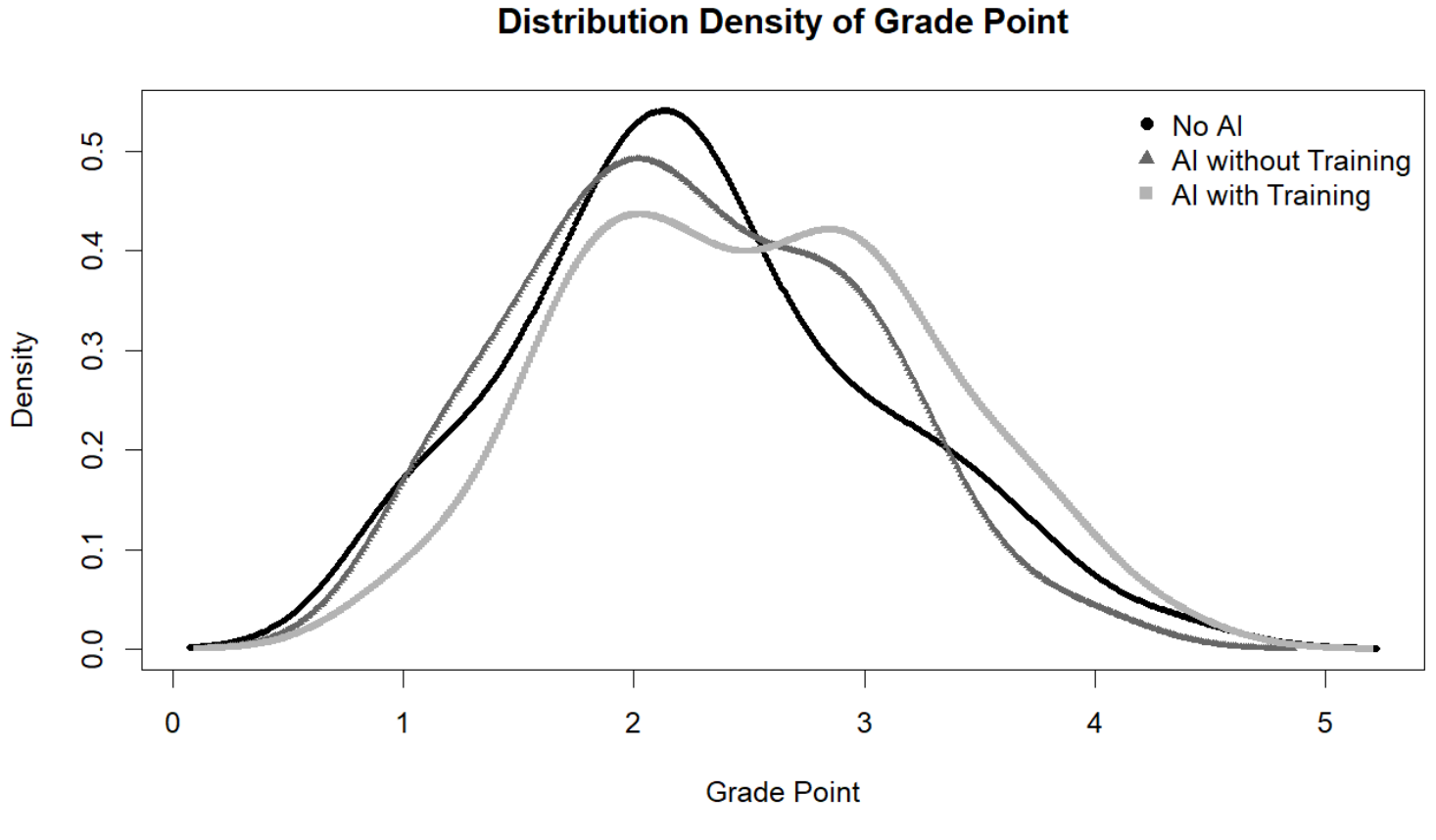

**Alt text:** Density plot titled "Distribution Density of Grade Point" showing three overlapping kernel density curves for Grade Point (x-axis, range 0–5) against Density (y-axis, range 0–0.55). Curves represent three groups: "No AI" (black solid line, peak at ~2.0, highest density), "AI without Training" (dark gray line, peak at ~2.0), and "AI with Training" (light gray line, peak at ~2.3 with a secondary rise near 3.0, showing a right-shifted and wider distribution than the other curves).

The first measure of performance is grade point. We hypothesized that access to DeepSeek without the training intervention would increase or decrease the average grade point in Group 2 compared to Group 1. We also hypothesized that access to DeepSeek with the training intervention would increase the average grade point in Group 3 compared to Groups 1 and 2.

**Figure 5** plots mean grade point by group. Subjects in Group 1 attained a mean grade point of 2.290 (SD = 0.779), subjects in Group 2 attained a mean grade point of 2.251 (SD = 0.714), and subjects in Group 3 attained a mean grade point of 2.521 (SD = 0.771). On average, participants in Group 2 seemed to perform worse than participants in Group 1 on the grade point measure but this decline is not statistically significant ($d = -0.039$; $p = 0.790$, two-tailed $t$-test). Participants in Group 3 did better than participants in Group 2 on the same measure, indicating that the treatment intervention increased the quality of answers ($d = 0.270$; $p = 0.027^{*}$, one-tailed $t$-test). Participants in Group 3 also appeared to outscore participants in Group 1 although the improvement does not attain conventional levels of statistical significance ($d = 0.231$; $p = 0.064$, one-tailed $t$-test).

**Figure 5:** Average Grade Point by Group

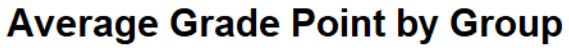

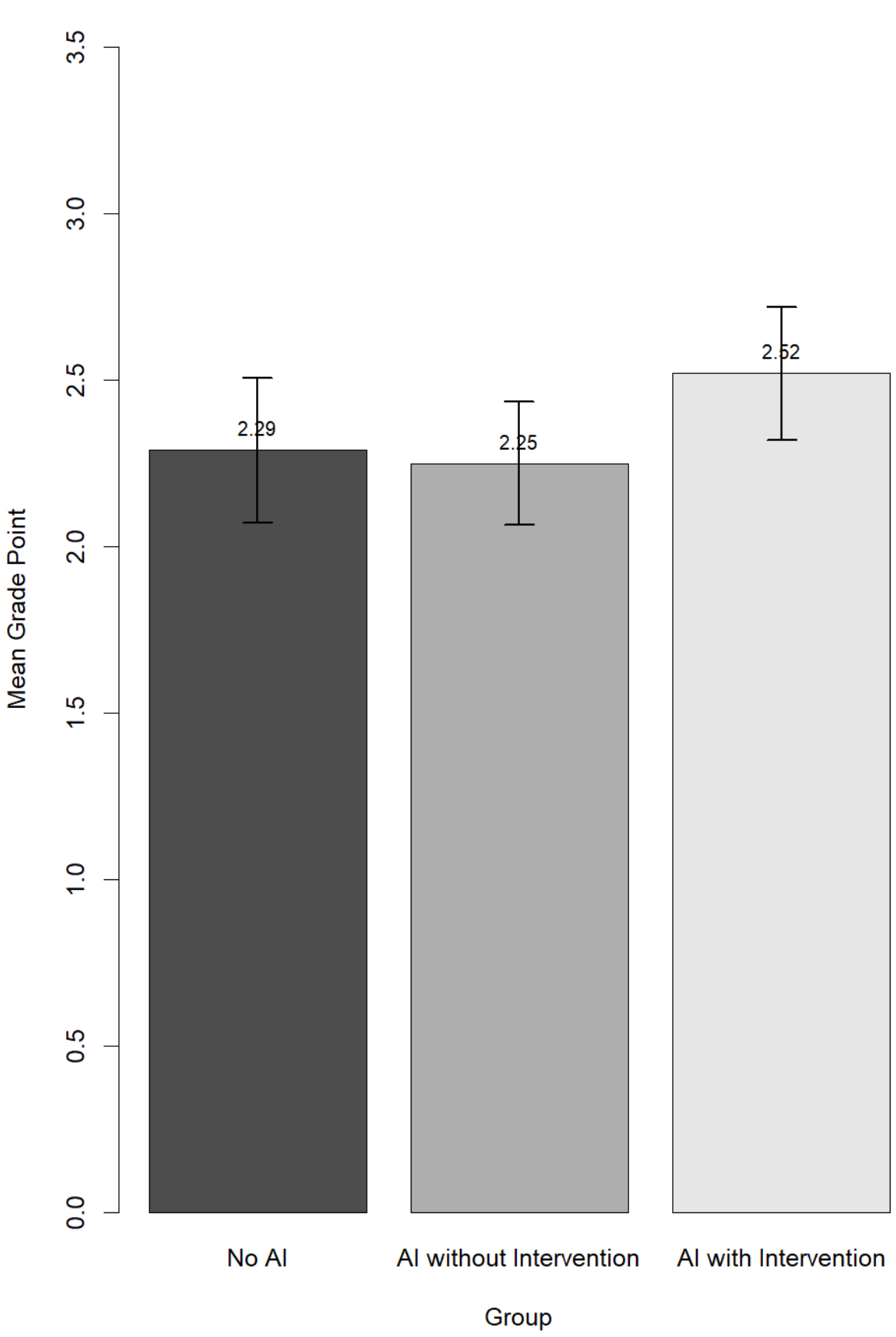

**Alt text:** Bar chart showing mean grade point by experimental group, with 95% confidence intervals. Group 1 (No AI, dark grey bar): 2.29; Group 2 (AI without Intervention, medium grey bar): 2.25; Group 3 (AI with Intervention, light grey bar): 2.52. The difference between Groups 2 and 3 is statistically significant ($d = 0.270$; $p = 0.027$*, one-tailed *t*-test). The difference between Groups 1 and 3 does not reach conventional significance ($d = 0.231$; $p = 0.064$).

The second measure of performance is the number of issues missed. We hypothesized that access to DeepSeek would decrease the average number of issues missed in Groups 2 and 3 compared to Group 1.

**Figure 6** plots mean number of issues missed by group. The mean number of issues missed is 1.102 (SD = 0.941) for subjects in Group 1, 1.228 (SD = 1.069) for subjects in Group 2 and 0.983 (SD = 0.927) for subjects in Group 3. These figures suggest, contrary to our expectations, that participants in Group 2 performed worse than participants in Group 1 on missed issues. This effect is, however, not statistically significant ($d = 0.126$; $p = 0.740$, one-tailed $t$-test). On average, participants in Group 3 missed fewer issues than participants in Group 1, but this difference is also not statistically significant ($d = -0.119$; $p = 0.256$, one-tailed $t$-test).

**Figure 6:** Average Number of Missed Issues by Group

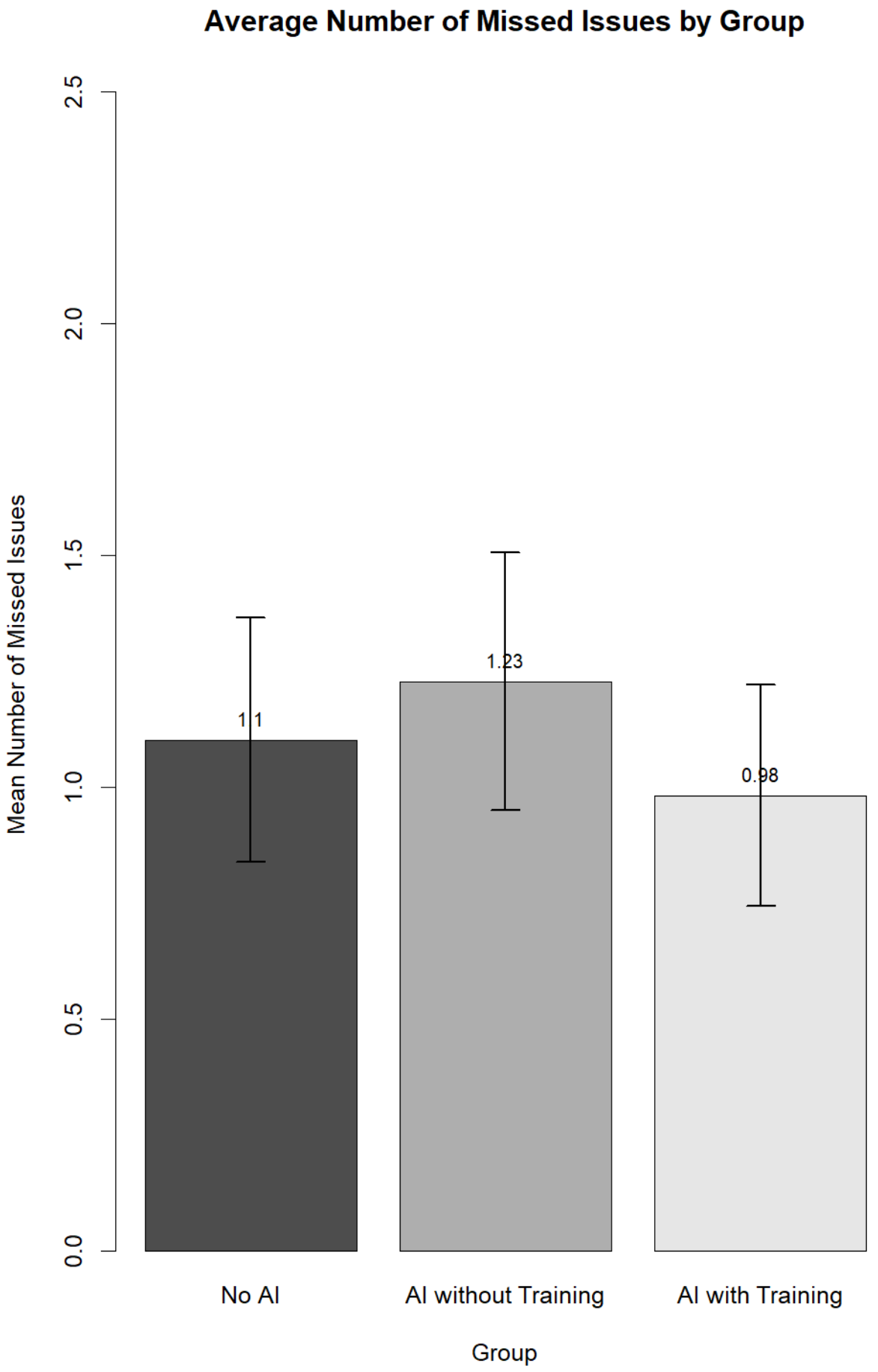

**Alt text**: Bar chart showing mean number of examination issues missed by group, with 95% confidence intervals. Group 1 (No AI, dark grey bar): 1.10; Group 2 (AI without Intervention, medium grey bar): 1.23; Group 3 (AI with Intervention, light grey bar): 0.98. No statistically significant differences are detected across groups on this measure.

Overall, there is no evidence that access to DeepSeek alone enhanced participants' performance on the examination as measured by grade point or by number of issues missed. The training intervention increased the proportion of participants who reported using DeepSeek on the examination. Among participants who had access to DeepSeek, the training intervention also enhanced performance as measured by grade point.

### 5.3. Comparing Answer Readability and Length

To get a sense of how access to DeepSeek and the training intervention may have modified subject behavior, we also examined the readability ("complexity") and length of answers. Readability was quantified using the Flesch–Kincaid grade level (Nielsen et al. 2024). We hypothesized that access to DeepSeek would increase or decrease the average complexity of answers in Groups 2 and 3 compared to Group 1. **Figure 7** plots the mean complexity of answers by group. The mean complexity is 12.845 (SD = 2.151) for Group 1, 12.379 (SD = 1.989) for Group 2 and 12.131 (SD = 1.911) for Group 3. There are no statistically significant differences in readability between answers in Group 1 and Group 2 ($d = -0.466$; $p = 0.253$, two-tailed $t$-test) or between Group 1 and Group 3 ($d = -0.714$; $p = 0.075$, two-tailed $t$-test). However, the slightly lower Flesch-Kincaid grade level for Group 2 and 3 may suggest the access to GenAI tools can help achieve more organized and less verbose writing.

**Figure 7:** Average Complexity by Group

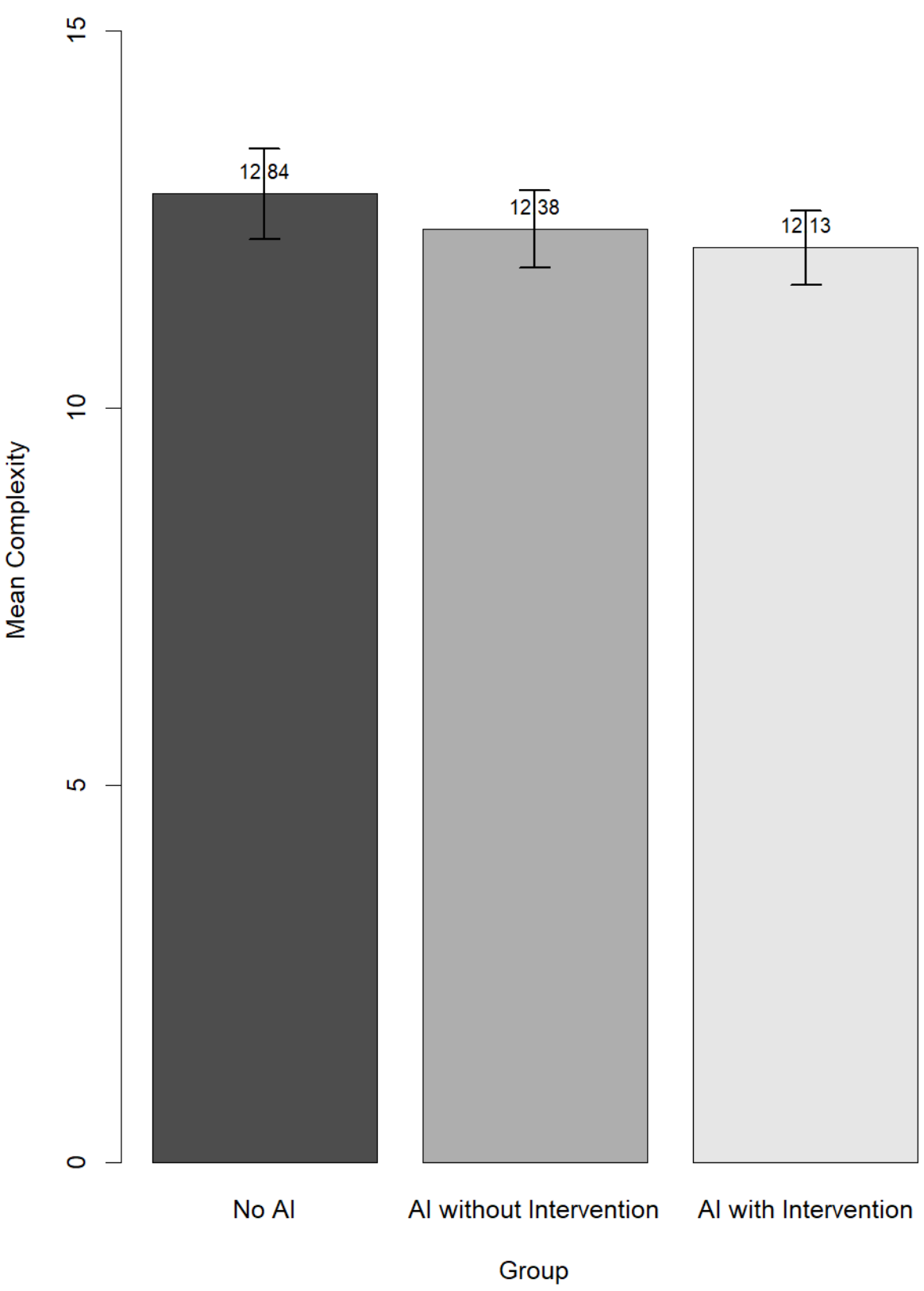

**Alt text:** Bar chart showing mean Flesch–Kincaid grade level of answers by group, with 95% confidence intervals, used as a measure of answer complexity. Group 1 (No AI, dark grey bar): 12.84; Group 2 (AI without Intervention, medium grey bar): 12.38; Group 3 (AI with Intervention, light grey bar): 12.13. No statistically significant differences are detected across groups.

Length of answers was quantified by total number of words of each answer. We hypothesized that access to DeepSeek would increase the average length of answers in Groups 2 and 3 compared to Group 1. **Figure 8** plots mean length of answers by group. The mean length of answers is 1059.673 (SD = 406.135) in Group 1, 895.737 (SD = 279.167) in Group 2, and 1001.638 in Group 3 (SD = 272.029). Contrary to our

expectations, answers in Group 2 ($d$ = -163.937; $p$ = 0.990, one-tailed $t$-test) and Group 3 ($d$ = -58.036; $p$ = 0.802, one-tailed $t$-test) were not longer, on average, than answers in Group 1. If anything, answers in Group 2 were on the whole shorter than answers in Group 1 ($d$ = -163.937; $p$ = 0.019*, two-tailed $t$-test) and Group 3 ($d$ = -105.901; $p$ = 0.042*, two-tailed $t$-test).[10]

**Figure 8:** Average Length by Group

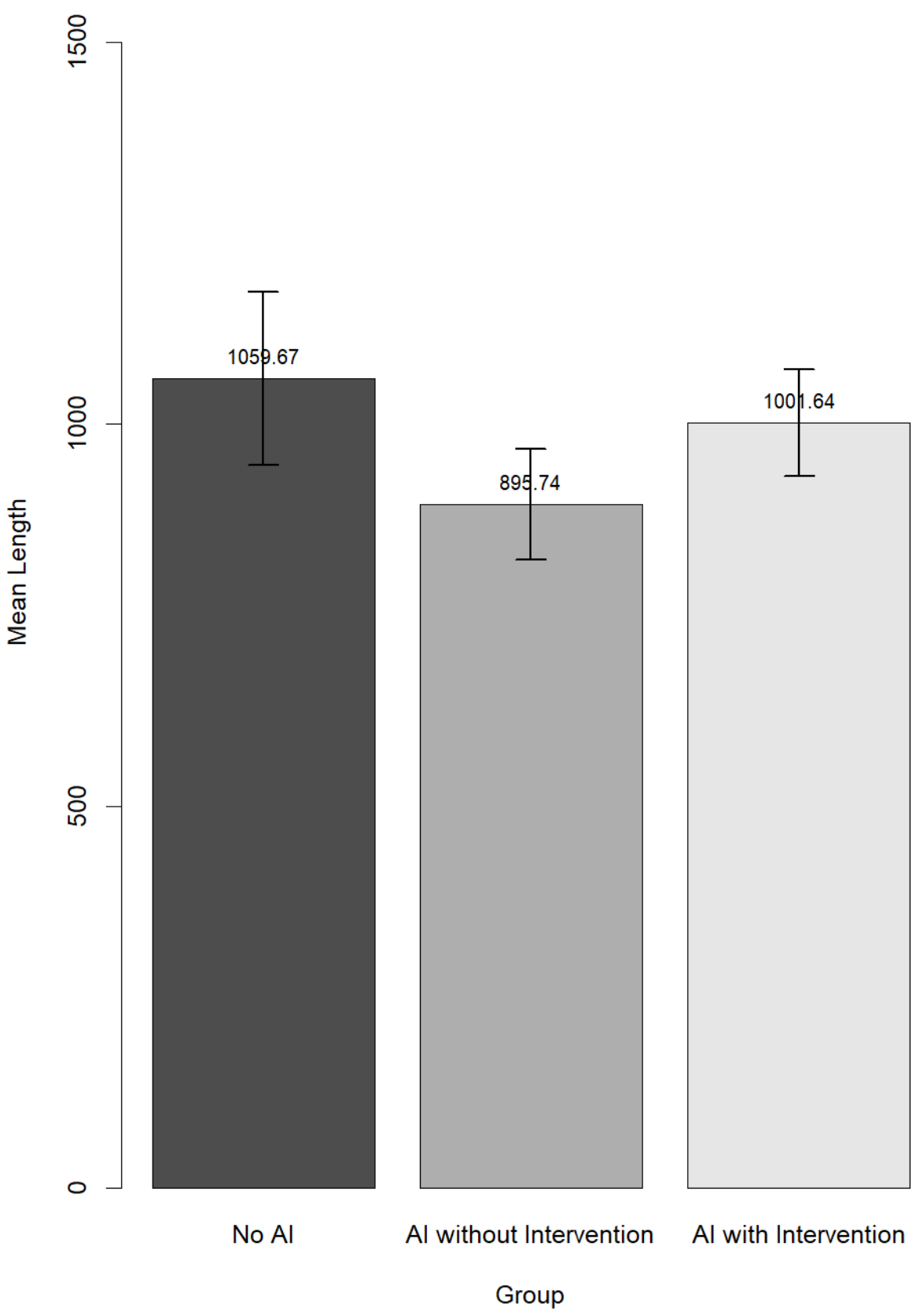

**Alt text:** Bar chart showing mean word count of answers by group, with 95% confidence intervals. Group 1 (No AI, dark grey bar): 1,059.67; Group 2 (AI without Intervention, medium grey bar): 895.74; Group 3

[10] These tests were not pre-registered.

(AI with Intervention, light grey bar): 1,001.64. Answers in Group 2 are significantly shorter than those in Groups 1 and 3 ($p = 0.019^*$ and $p = 0.042^*$, respectively, two-tailed t-tests; tests not pre-registered).

These reversals are intriguing because it seems almost obvious that answers should be longer when drafted with the help of GenAI. One possible explanation for our data is that Group 2 participants who had access to DeepSeek without receiving the training intervention spent time interacting unfruitfully with the LLM.

### 5.4. Comparing Authority Citation Accuracy

We also assessed the accuracy of the authorities cited by participants. Three key variables were analyzed: Rule Misstatements (number of erroneous statements or omissions of rules), Case Hallucinations (number of non-existing cases), and Case Misstatements (number of cases irrelevant to the preceding argument). The examination included four major issues; each associated with one point for the "accurate statement of the applicable rule for that issue." The Rule Misstatements variable was calculated by counting instances of rule omissions or misstatements across these four issues. Thus, participants could score between 0 and 4 on this variable: a score of 0 indicated accurate statement of all rules, while a score of 4 reflected erroneous statement or omission of all applicable rules. For Case Hallucinations, one point was assigned for each cited case that does not exist in real-world legal or regulatory databases. For Case Misstatements, one point was assigned for each cited case that fails to support or is irrelevant to the participant's preceding argument.

We hypothesized that access to DeepSeek alone would increase the numbers of all three variables in Group 2 compared to Group 1. We also hypothesized that the training intervention would decrease the numbers of all three variables in Group 3 compared to Group 2. We recorded zero Case Hallucination for all the 164 answers. This finding is within expectation, as participants cited relatively few cases (Average = 5.493) during the one-hour exam. This may also be attributed to their familiarity with case names from course materials, which equipped them with basic judgment to avoid referencing non-existent cases.

**Figure 9** shows the mean number of Rule Misstatements is 2.469 (SD = 1.192) for participants in Group 1, 2.579 (SD = 0.944) for participants in Group 2 and 2.155 (SD = 1.105) for participants in Group 3. Group 2 recorded a significantly higher number of Rule Misstatements compared with Group 3 ($d = -0.424$; $p = 0.014$*, one-tailed $t$-test), which suggests training can effectively help participants to state rule more accurately. There are no statistically significant effects between Group 2 and Group 1 ($d = 0.110$; $p = 0.303$, one-tailed $t$-test).

**Figure 9:** Average Number of Rule Misstatements by Group

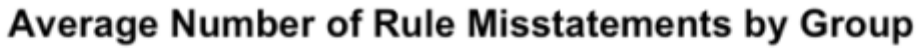

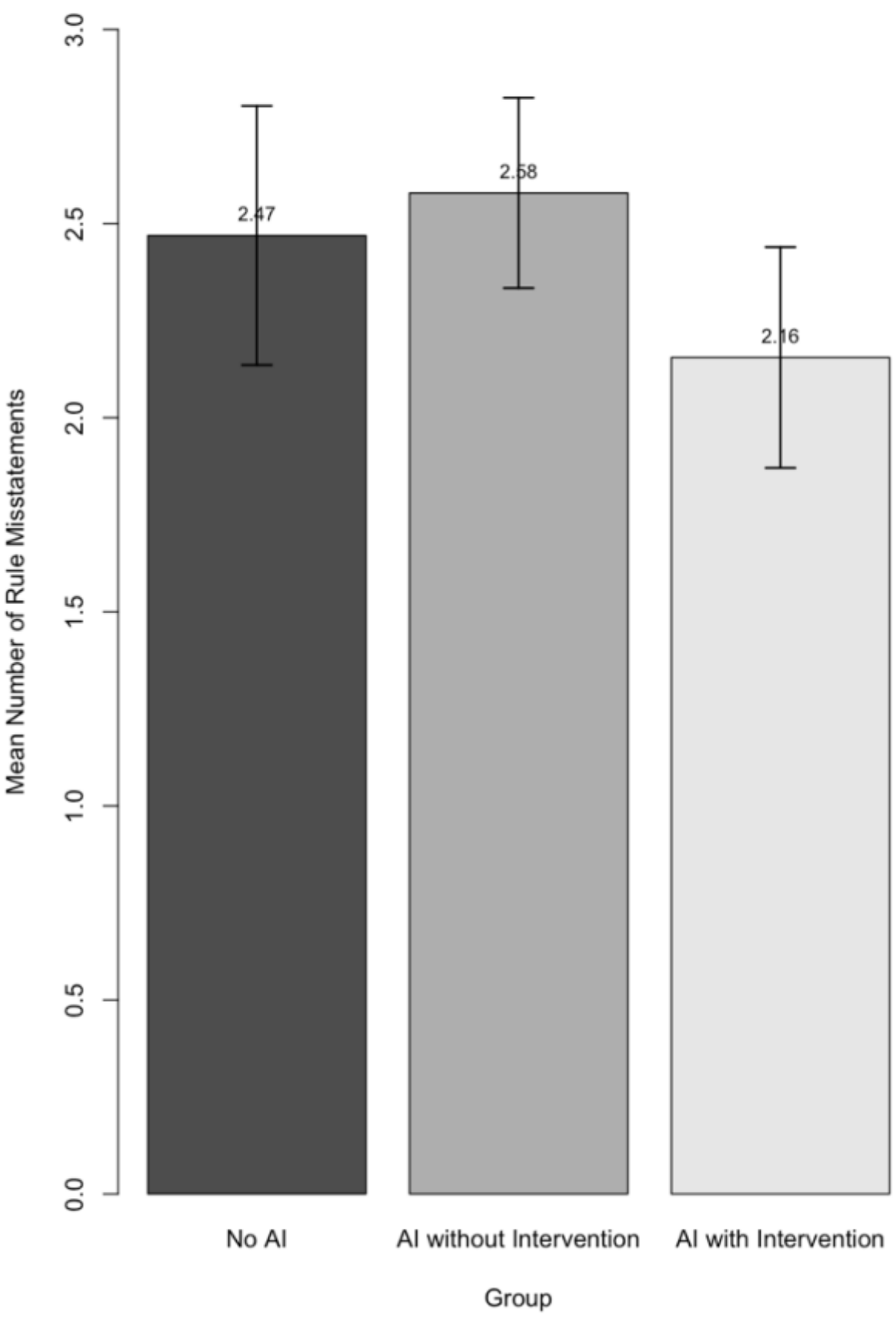

**Alt text:** Bar chart showing mean number of rule misstatements by group, with 95% confidence intervals. Group 1 (No AI, dark grey bar): 2.47; Group 2 (AI without Intervention, medium grey bar): 2.58; Group 3 (AI with Intervention, light grey bar): 2.16. The difference between Groups 2 and 3 is statistically significant ($d = -0.424$; $p = 0.014$*, one-tailed $t$-test), indicating that training reduced inaccurate or incomplete rule statements.

**Figure 10** shows the mean number of Case Misstatements is 0.020 (SD = 0.143) for participants in Group 1, 0.193 (SD = 0.581) for participants in Group 2 and 0.207 (SD = 0.614) for participants in Group 3. Group 2 recorded a significantly higher number of Case Misstatements compared with Group 1 ($d = 0.173$; $p = 0.017$*, one-tailed $t$-test). Considering there is also a significantly smaller number of total cases cited by Group 2 compared with Group 1 ($d$ = -1.520; $p$ = 0.017*, two-tailed $t$-test[11]), a possible explanation is that the ineffective interaction with GenAI tools by Group 2 ultimately left participants with less time to write their answers (echoes the discussion of length), to refer to authorities and to verify authorities. There are no statistically significant effects between Group 2 and Group 3 ($d$ = 0.014; $p$ = 0.550, one-tailed $t$-test).

**Figure 10:** Average Number of Case Misstatements by Group

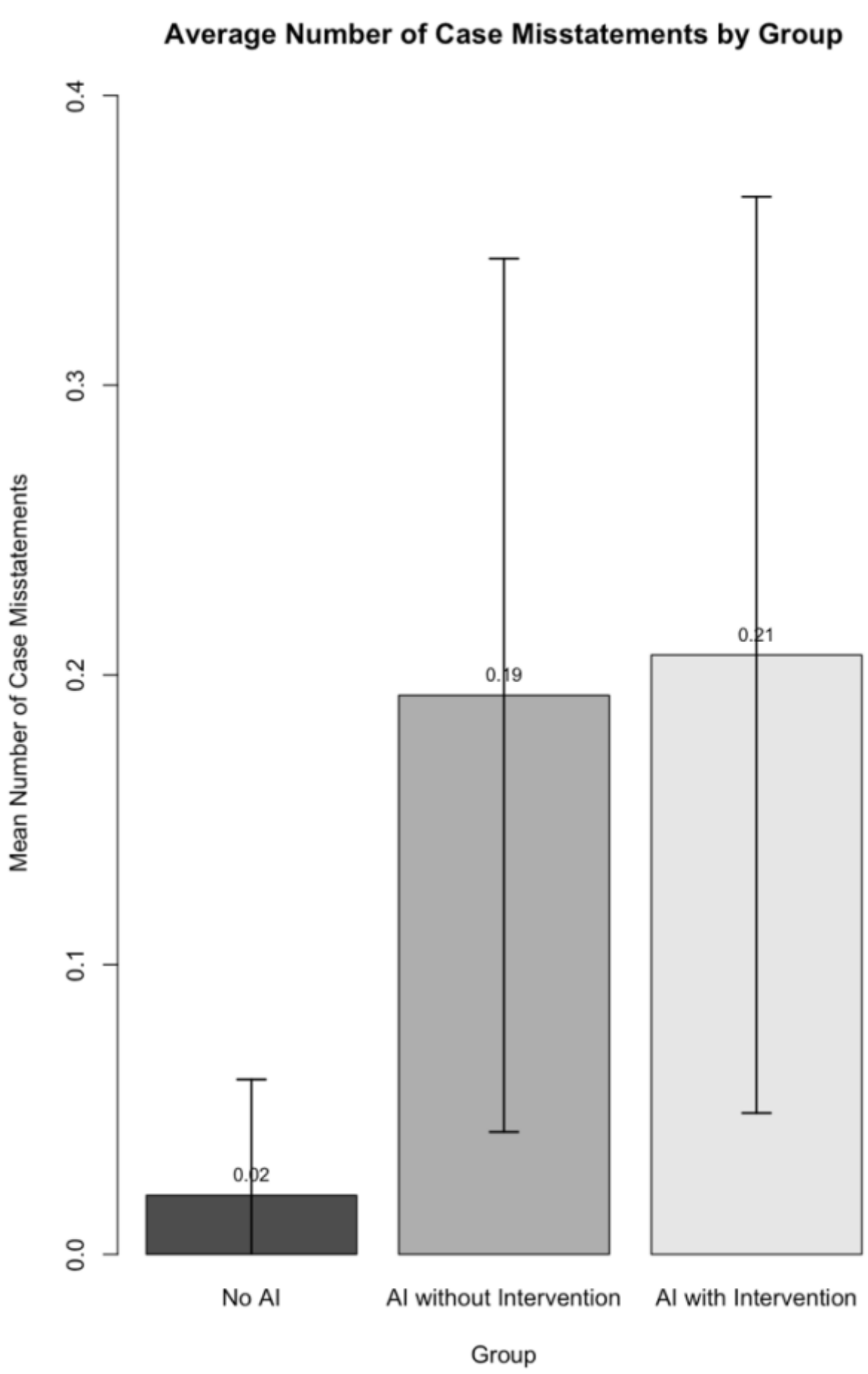

**Alt text:** Bar chart showing mean number of case misstatements by group, with 95% confidence intervals. Group 1 (No AI, dark grey bar): 0.02; Group 2 (AI without Intervention, medium grey bar): 0.19; Group 3

[11] This test was not pre-registered.

(AI with Intervention, light grey bar): 0.21. Group 2 has significantly more case misstatements than Group 1 ($d$ = 0.173; $p$ = 0.017*, one-tailed $t$-test). The difference between Groups 2 and 3 is not statistically significant.

### 5.5. Decomposing the Effect of Training on Performance

Overall, our intervention increased self-reported use of DeepSeek and improved performance on the mock examination for participants in Group 3. Did training raise the average grade point by enhancing the effectiveness of DeepSeek use for those who would have used it anyway or by encouraging the adoption of DeepSeek by those who would not have used it otherwise?

The standard approach to evaluating training interventions uses instrumental variables (IV) to estimate the Local Average Treatment Effect (LATE) of treatment on outcomes (Angrist et al. 1996). This approach treats assignment to treatment as an instrument for treatment and estimates the effect of adoption for compliers, i.e. those who take the treatment because of assignment to treatment. However, this framework rests critically on an exclusion restriction: that the instrument affects outcomes only through its effect on the endogenous variable, e.g. individuals choosing to take treatment. This assumption is not satisfied in many realistic settings. Here, training may impact outcomes through two distinct channels. First, by inducing non-adopters to adopt GenAI—the adoption effect, and second, by changing how existing users employ GenAI—the effectiveness effect. When the exclusion restriction fails, the LATE conflates both mechanisms.

We address this identification challenge using principal stratification (Frangakis and Rubin 2002) with partial identification. Following the potential outcomes framework (Rubin 1974, 1976, 1997; Splawa-Neyman et al. 1990), let $Z \in \{0, 1\}$ represent assignment to the training intervention, $D(Z) \in \{0,1\}$ represent potential use of DeepSeek given assignment to training, and $Y(Z, D)$ represent potential grade point given DeepSeek use and assignment to training. Never users, $N$, are participants for whom $D(0) = D(1) = 0$. Always users, $A$, are participants for whom $D(0) = D(1) = 1$. Induced users, $C$, are participants for whom $D(0) = 0$ and $D(1) = 1$. Defiers are participants for whom $D(0) = 1$ and $D(1) = 0$.

Define $\tau_{adoption} \equiv E[Y(1,1) - Y(0,0)|C]$ as the adoption effect for induced users and $\tau_{effectiveness} \equiv E[Y(1,1) - Y(0,1)|A]$ as the effectiveness effect for always users. We derive sharp bounds on the adoption and effectiveness effects under three plausible assumptions:

1) monotone treatment response, i.e. $D(1) \geq D(0)$ for all participants;
2) an exclusion restriction for never users, i.e. $Y(0,1) = Y(0,0)$; and
3) a global support restriction that grade points must lie in [$Y_{min} = 1.0$, $Y_{max} = 4.3$] for all participants under all potential outcome states.

The monotone treatment response assumption posits that training can only encourage, not discourage, DeepSeek use and that there are therefore no defiers. The exclusion restriction for never users claims that training has no effect on participants who would not use DeepSeek, even if they received the intervention.

Several quantities are point-identified under random assignment and monotone treatment response, as stated in the following propositions.

**Proposition 1 (Stratum Proportions).** Under random assignment and monotone treatment response:

i. $\pi_A = P(D = 1|Z = 0)$

ii. $\pi_N = P(D = 0|Z = 1)$

iii. $\pi_C = P(D = 0|Z = 0) - P(D = 0|Z = 1)$

**Proposition 2 (Baseline Potential Outcomes).** Under monotonicity and the exclusion restriction for never users:

i. $E[Y(0,0)|N] = E[Y|Z = 1, D = 0]$

ii. $E[Y(0,1)|A] = E[Y|Z = 0, D = 1]$

iii. $E[Y(0,0)|C] = \frac{E[Y|Z = 0, D = 0] - w_N \times E[Y(0,0)|N]}{w_C}$

Trained users here are a mixture of induced and always users:

$$E[Y|Z = 1, D = 1] = w_C \times E[Y(1,1)|C] + w_A \times E[Y(1,1)|A]$$

where $w_C = \frac{\pi_C}{\pi_C + \pi_A}$ and $w_A = \frac{\pi_A}{\pi_C + \pi_A}$. $E[Y(1,1)|C]$ and $E[Y(1,1)|A]$ are not directly observed. Hence, unless additional assumptions are made, both adoption and effectiveness effects are partially identified.

**Proposition 3 (Bounds Under Global Support Restriction).** Under monotone treatment response and the global support restriction,

$$\tau_{adoption} \in [L_{adoption}, Y_{max} - E[Y(0,0)|C]]$$

where

$$L_{adoption} = \frac{(E[Y|Z = 1, D = 1] - w_A \times Y_{max})}{w_C} - E[Y(0,0)|C]$$

and

$$\tau_{effectiveness} \in [Y_{min} - E[Y(0,1)|A], U_{effectiveness}]$$

where

$$U_{effectiveness} = \frac{(E[Y|Z = 1, D = 1] - w_C \times Y_{min})}{w_A} - E[Y(0,1)|A].$$

**Table 2** reports empirical bounds under global support restriction. The bounds are wide and uninformative about the sign of either effect.

**Table 2:** Bounds Under Support Restriction Only

| *Effect* | *Lower Bound* | *Upper Bound* | *Width* |
|---|---|---|---|
| Adoption | -0.454 | 2.846 | 3.300 |
| Effectiveness | -1.093 | 1.226 | 2.319 |

**Alt Text:** Sharp bounds on adoption and effectiveness effects under the support restriction alone. The adoption effect is E[Y(1,1) − Y(0,0) | Induced Users] and the effectiveness effect is E[Y(1,1) − Y(0,0) | Always Users]. Bounds are wide and uninformative about the sign of either effect.

To tighten the bounds, we introduce another assumption:

4) mean dominance, i.e. $E(Y(1,1)|C) \geq E(Y(1,1)|A) - \gamma$ for $\gamma \geq 0$.

This assumption states that trained induced users perform as well as trained always users up to a tolerance $\gamma$. The parameter $\gamma$ allows relaxation of strict dominance for sensitivity analysis.

**Proposition 4 (Bounds Under Mean Dominance).** Under monotone treatment response, the global support restriction, and mean dominance, define the upper bound on $E[Y(1,1)|A]$ as

$$U_{1,1,always} = \min\{E[Y|Z=1, D=1] + w_C \times \gamma, Y_{max}\}$$

and the lower bound on $E[Y(1,1)|C]$ as

$$L_{1,1,induced} = \max\left\{\frac{E[Y|Z=1, D=1] - w_A \times U_{1,1,always}}{w_C}, Y_{min}\right\}.$$

Then

$$\tau_{adoption} \in \left[L_{1,1,induced} - E[Y(0,0)|C], Y_{max} - E[Y(0,0)|C]\right]$$

and

$$\tau_{effectiveness} \in \left[Y_{min} - E[Y(0,1)|A], U_{1,1,always} - E[Y(0,1)|A]\right].$$

**Table 3** reports empirical bounds under mean dominance for varying levels of $\gamma$. It also reports 95% confidence intervals constructed with 2000 replications using the percentile bootstrap method (Imbens and Manski 2004).

**Table 3:** Bounds as Functions of Mean Dominance Relaxation

<table>
<tr><th>$\gamma$</th><th>$L_{effectiveness}$</th><th>$U_{effectiveness}$</th><th>$L_{adoption}$</th><th>$U_{adoption}$</th></tr>
<tr><td>0.000</td><td rowspan="5">-1.093<br>[-1.473, -0.702]</td><td>0.419<br>[-0.049, 0.938]</td><td>1.059<br>[-0.364, 1.740]</td><td rowspan="5">2.846<br>[ 1.449, 3.300]</td></tr>
<tr><td>0.400</td><td>0.565<br>[ 0.083, 1.096]</td><td>0.805<br>[-0.627, 1.472]</td></tr>
<tr><td>0.640</td><td>0.652<br>[ 0.124, 1.207]</td><td>0.652<br>[-0.786, 1.322]</td></tr>
<tr><td>1.000</td><td>0.783<br>[ 0.215, 1.400]</td><td>0.423<br>[-1.009, 1.115]</td></tr>
<tr><td>1.665</td><td>1.025<br>[ 0.325, 1.742]</td><td>0.000<br>[-1.455, 0.710]</td></tr>
</table>

**Alt Text:** Sharp bounds on adoption and effectiveness effects as functions of the mean dominance relaxation parameter $\gamma$, with 95% bootstrap confidence intervals (2,000 replications) in brackets. Mean dominance requires E[Y(1,1) | Induced Users] $\geqslant$ E[Y(1,1) | Always Users] − $\gamma$, with $\gamma = 0$ corresponding to strict dominance. The effectiveness lower bound (−1.093) and the adoption upper bound (2.846) are constant across all values of $\gamma$ and determined by the support restriction alone. The crossover point $\gamma$ = 0.640 marks the value at which the adoption lower bound equals the effectiveness upper bound.

Point estimates suggest the adoption effect for induced users exceeds the effectiveness effect for always users: at $\gamma = 0$, the adoption lower bound exceeds the effectiveness upper bound by 0.64 grade points. Confidence intervals are wide and overlapping, so we cannot statistically reject effectiveness as the dominant mechanism, but the point estimate ordering holds unless always users outperform induced users post-training by more than this margin.

## 6. Discussion

We distinguished at the outset between two channels through which generative artificial intelligence may affect professional productivity: the scope of use—whether users choose to adopt the technology for a given task—and the effectiveness of use conditional on adoption. The experimental evidence presented here indicates that user training plays an important role along the extensive margin. Training significantly increases the likelihood that participants employ GenAI for boundary applications, expanding the scope of use to tasks that would otherwise be avoided. Moreover, GenAI adoption induced by training is associated with higher-quality legal output, suggesting that adoption in these settings is productive when accompanied by appropriate instruction.

At the same time, the design of our experiment does not permit sharp identification of whether training also improves the effectiveness of use among individuals who would adopt GenAI regardless of instruction. Our principal stratification analysis nevertheless indicates that improvements in overall performance may plausibly reflect a combination of expanded adoption and more effective use. Conversely, access to GenAI without training does not improve performance on average and may be unproductive, although the available evidence does not allow firm conclusions about negative effects. Taken together, these findings underscore that the productivity consequences of GenAI depend critically on how and by whom the technology is used.

These results carry direct implications for legal education and for the organization of legal practice. In educational settings, they support proposals to incorporate instruction on effective human–AI collaboration as a core component of professional training. Such instruction should emphasize both practical skills—such as how to structure prompts and interrogate model output—and conceptual understanding of how GenAI systems function and where they are prone to error. For law firms and legal departments, the findings suggest that successful GenAI integration requires coordinated investment in both technology access and human capital. Providing access to GenAI tools without accompanying training may yield little benefit and could even undermine short-run productivity, potentially fostering resistance to adoption that slows longer-run innovation.

More broadly, our findings contribute to debates in labor economics on task-based technological change, the evolution of professional work, and the returns to training investment. Traditional models of skill-biased and task-based technological change treat cognitive non-routine tasks—typically performed by high-skill workers—as relatively insulated from productivity-enhancing automation (Acemoglu and Autor 2011; Acemoglu and Restrepo 2019; Autor et al. 2003).

Yet, human capital investments can determine patterns of GenAI adoption. Our evidence suggests that higher-ability individuals tend to be induced users, adopting GenAI only when training reduces perceived risks, whereas lower-ability individuals are more likely to adopt regardless. The data are consistent with this interpretation, though with an important caveat: because grade point is itself a post-treatment outcome from the same examination, it cannot serve as a clean pre-treatment measure of ability. The quartile classification is therefore potentially endogenous to the treatment, and the pattern in Table 4 should be interpreted as descriptive rather than confirmatory. Adopters in Group 2 had lower grade points than non-adopters. The introduction of training in Group 3 leveled adoption across the ability distribution, with the most dramatic increases occurring among those in the top-grade point quartile (**Table 4**).

**Table 4:** DeepSeek Adoption Rates by Grade Point Quartile and Group

| Grade Point Quartile | Group 2 Rate | Group 3 Rate |
|---|---|---|
| Q1 ( ≤ 1.70) | 33.3% (6/18) | 38.5% (5/13) |
| Q2 (1.70 - 2.30) | 16.7% (3/18) | 43.8% (7/16) |
| Q3 (2.30 - 3.00) | 40.0% (6/15) | 41.2% (7/17) |
| Q4 ( > 3.00) | 0.0% (0/6) | 41.7% (5/12) |

**Alt Text:** DeepSeek adoption rates by grade point quartile and experimental group. Quartile thresholds are based on grade points pooled across Groups 2 and 3. Numerators and denominators in parentheses indicate users over total participants in each cell. Training substantially increases adoption in the upper quartile (from 0.0% in Group 2 to 41.7% in Group 3), consistent with the prediction that higher-ability individuals are induced users.

These results undermine the narrative that GenAI primarily benefits lower-skilled workers. In the absence of training, less-skilled individuals may adopt the technology but use it ineffectively, while more-skilled individuals opt out. When training is available, however, adoption rises among higher-ability professionals, unlocking productivity gains in judgment-intensive tasks that have traditionally resisted technological substitution.

Finally, the results speak to the diffusion of GenAI as a general-purpose technology (Bresnahan and Trajtenberg 1995). While GenAI exhibits broad applicability and the potential for complementary innovation, our findings indicate that realizing these benefits requires learning and adaptation that are often task-specific. The brief training intervention in our study focused narrowly on legal issue-spotting rather than general AI literacy. That even such minimal, application-focused instruction improves outcomes is encouraging for diffusion. At the same time, the task-specific nature of effective training implies that different occupations—and even individual organizations—may need to develop tailored training protocols. Such customization may slow diffusion and generate temporary advantages for early adopters who successfully integrate GenAI into their workflows.

## 7. Limitations

Several limitations qualify our conclusions. First, the experimental setting—an issue-spotting examination—captures only a subset of professional legal practice. While such examinations assess individual legal reasoning under time constraints, legal work in practice is often collaborative, iterative, and embedded in organizational processes. In addition, our sample consists of law students from a single institution. The effects of user training on GenAI adoption and performance may differ across levels of professional experience, practice areas, or institutional environments. These considerations limit external validity and suggest the value of replications in more naturalistic or field-based settings.

Second, our performance measures reflect a discrete work product produced over a short time horizon. Although examination scores provide a meaningful and widely used proxy for legal competence, they do not reveal whether the benefits of training persist or evolve with repeated use. Longitudinal studies that track behavior and performance over extended periods would be useful for assessing the durability of training interventions and potential learning or decay effects.

Third, our principal stratification analysis relies on assumptions—including monotonicity and mean dominance—that are standard in the econometrics literature but not directly testable. While these assumptions are consistent with our experimental design and observed data, violations could affect the interpretation of the bounds we derive. Future research could explore alternative identification strategies and more rigorous ability measurement. Our use of exam GPA as a post-hoc proxy for ability is suggestive but imperfect: the quartile cell sizes are small and the analysis was not pre-registered. A pre-registered design with an independent pre-treatment ability measure would permit a more rigorous test of the prediction that training disproportionately induces adoption among higher-ability professionals.

Finally, our analysis focuses on individual-level performance and abstracts from organizational and institutional factors that shape technology adoption in professional settings. In practice, law firms face collective action problems in training investment, uncertainty about the quality and returns to training, and agency issues between firm management and individual attorneys. Understanding how organizational structure and incentives influence GenAI training and adoption represents an important direction for future research.

More broadly, our single intervention in a controlled setting leaves several questions open. How do training effects vary across different GenAI tools and professional domains? What are the optimal timing, intensity, and content of training interventions? How do organizational and labor market institutions shape access to training and patterns of adoption? And how do the returns to training evolve as GenAI technologies improve? Addressing these questions will require research programs that combine experimental, observational, and computational approaches across diverse contexts.

## 8. Conclusion

The emergence of generative artificial intelligence parallels earlier technological transformations that reshaped knowledge-intensive occupations, but it differs in a fundamental respect. Whereas computers and the internet primarily enhanced the storage, retrieval, and transmission of information, large language models attempt to perform analytical and generative functions that have traditionally defined professional expertise. This shift brings GenAI closer to the core of professional work than prior technologies, heightening both its potential value and its associated risks.

These features create a central tension in GenAI adoption. On the one hand, the high opportunity cost of professional time creates strong incentives to deploy tools that promise efficiency gains in research, drafting, and analysis. On the other hand, professional accountability and the high cost of error constrain reliance on technologies whose outputs cannot be taken at face value. LLMs embody this tension acutely: they can accelerate legal work while also producing confident but fabricated citations, misstatements of legal doctrine, and reasoning that appears plausible yet is fundamentally flawed. Whether GenAI enhances or undermines professional performance therefore depends critically on how it is used.

This paper advances a simple theoretical proposition: user training can shape the productivity effects of GenAI through two channels—by expanding the scope of use to tasks for which professionals would otherwise refrain from adoption, and by improving the effectiveness of use conditional on adoption. Evidence from our study shows that even a brief training intervention increases GenAI adoption for boundary applications and improves the quality of legal analysis produced, from comprehensively spotting issues to accurately stating rules, under time constraints. The results further provide suggestive, though not definitive, evidence that training may also enhance the returns to GenAI use among individuals who would adopt the technology even in the absence of instruction.

For legal educators and practitioners, these findings underscore the importance of coordinated investment in both GenAI tools and user training. Access to GenAI alone is unlikely to deliver reliable productivity gains and may yield limited or inconsistent returns. By contrast, even modest investments in training can meaningfully improve performance on tasks that are central to legal practice. More broadly, the results contribute to ongoing debates about skill-biased technological change, human capital investment, and the diffusion of general-purpose technologies. The productivity effects of GenAI are not technologically predetermined but depend on complementary investments in knowledge and capability. Moreover, the task-specific nature of effective GenAI use suggests that productivity gains may emerge unevenly over time, as organizations experiment with, refine, and institutionalize training practices that enable productive human–AI collaboration.

# Appendix

## Comparing Participant Preferences and Perceptions

We further examined the attitudes of participants toward GenAI as a tool for legal analysis and found no significant results. The two variables here are preferences about allowing GenAI to be used in law examinations, i.e., permission; and perceptions about the helpfulness of using GenAI for law examinations, i.e., helpfulness. Both variables were elicited on a 1 to 5 scale, with a higher score indicating a stronger preference for permission or a stronger perception of helpfulness.

We hypothesized that access to DeepSeek would increase or decrease the average permission score in Groups 2 or 3 compared to Group 1. We also hypothesized that the training intervention would increase the average permission score in Group 3 compared to Group 2.

**Figure 11** shows the mean permission score is 3.571 (SD = 1.118) for participants in Group 1, 3.228 (SD=1.086) for participants in Group 2 and 3.379 (SD=1.167) for participants in Group 3. There are no statistically significant effects between Group 1 and Group 2 ($d$ = -0.343; $p$ = 0.113, two-tailed $t$-test), between Group 1 and Group 3 ($d$ = -0.192; $p$ = 0.387, two-tailed $t$-test), or between Group 2 and Group 3 ($d$ = 0.151; $p$ = 0.237, one-tailed $t$-test).

**Figure 11:** Average Permission by Group

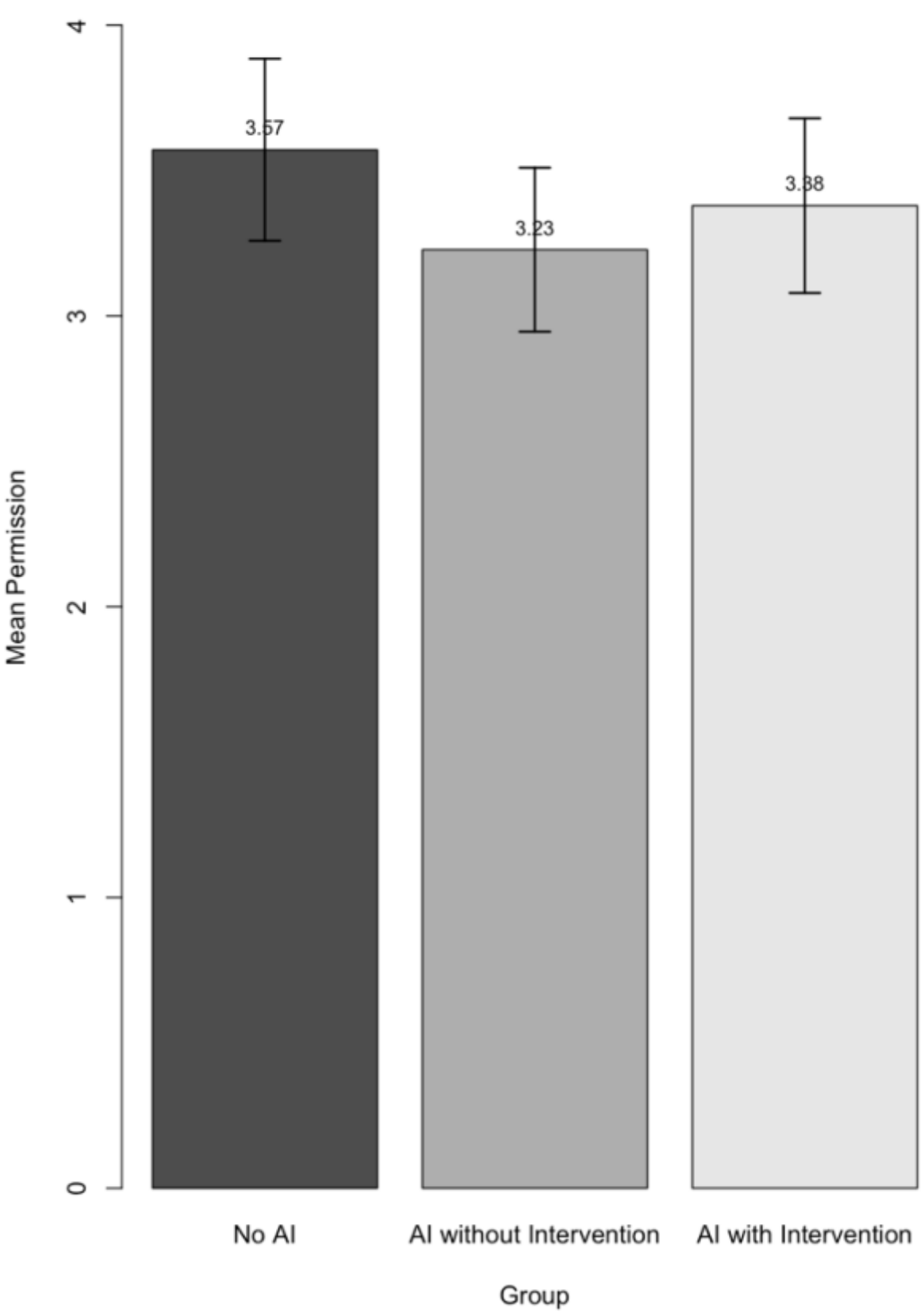

**Alt text**: Bar chart showing mean permission score by group, with 95% confidence intervals. Participants rated their agreement with allowing LLM access in law examinations on a 1–5 scale. Group 1 (No AI, dark grey bar): 3.57; Group 2 (AI without Intervention, medium grey bar): 3.23; Group 3 (AI with Intervention, light grey bar): 3.38. No statistically significant differences are detected across groups.

The lack of variation between the groups after some subjects had been given the opportunity and even training to use DeepSeek for an issue-spotter may not be entirely surprising. Given that examination grades are often awarded on a curve, participants could have taken into account not just the absolute improvement in their performance from using GenAI but also their improvement relative to others who also have access to the same technology. In this regard, we should note that the average permission scores for users in Group 2 ($d = 0.414$, $p = 0.290$; two-tailed $t$-test) and Group 3 ($d = 0.206$, $p = 0.522$; two-tailed $t$-test) are higher but ultimately not significantly different from the mean scores for non-users in these groups.[12]

Turning to perception, we hypothesized that access to DeepSeek without the training intervention would decrease the average score in Group 2 compared to Group 1. We also hypothesized that access to DeepSeek with the training intervention would increase or decrease the average score in Group 3 compared to Groups 1 and 2. **Figure 12** shows the mean helpfulness score is 3.122 (SD = 1.269) for participants in Group 1, 3.246 (SD = 1.405) for participants in Group 2, and 3.293 (SD = 1.298) for participants in Group 3. Again, there are no statistically significant effects between Group 1 and Group 2 ($d = 0.123$; $p = 0.682$, one-tailed $t$-test), between Group 1 and Group 3 ($d = 0.171$; $p = 0.494$, two-tailed $t$-test), or between Group 2 and Group 3 ($d = 0.047$; $p = 0.851$, two-tailed $t$-test). At the same time, users in Groups 2 and 3 rated the helpfulness of GenAI more highly than non-users in the same group. The difference for Group 2 is not statistically significant ($d = 0.571$, p = 0.186; two-sided $t$-test) while the difference for Group 3 is ($d = 0.922$, p = $0.005^{**}$; two-sided $t$-test).[13]

---

[12] These tests were not pre-registered.

[13] These tests were not pre-registered.

**Figure 12:** Average Helpfulness by Group

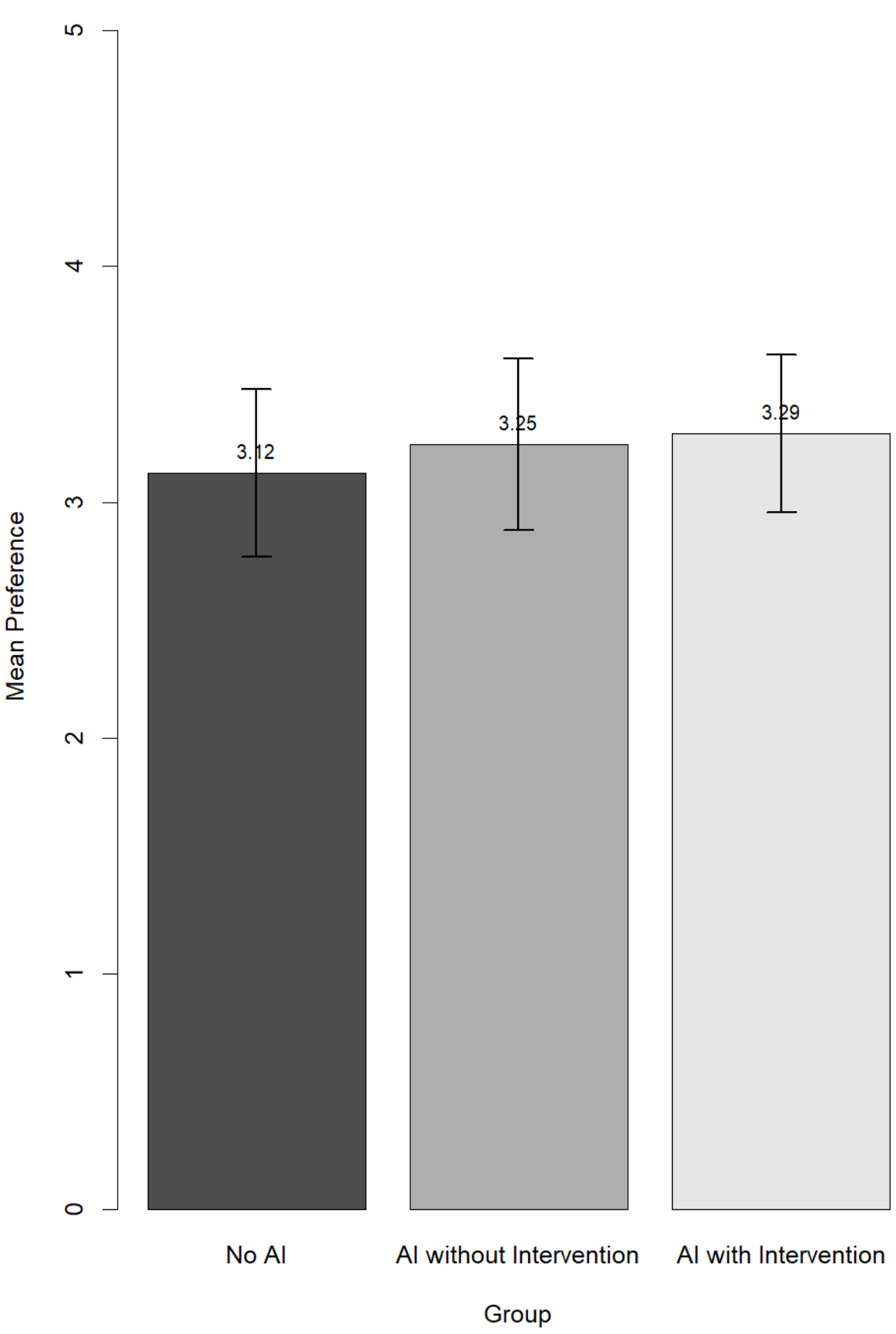

**Alt text:** Bar chart showing mean helpfulness score by group, with 95% confidence intervals. Participants rated their agreement that LLM access would help them in law examinations on a 1–5 scale. Group 1 (No AI, dark grey bar): 3.12; Group 2 (AI without Intervention, medium grey bar): 3.25; Group 3 (AI with Intervention, light grey bar): 3.29. No statistically significant differences are detected across groups, though DeepSeek users in Group 3 rated helpfulness significantly higher than non-users in the same group ($d$ = 0.922; $p$ = 0.005**, two-tailed $t$-test; not pre-registered).

## References

Acemoglu, Daron. 2025. "The Simple Macroeconomics of AI." *Economic Policy* 40 (121): 13–58. https://doi.org/10.1093/epolic/eiae042.

Acemoglu, Daron, and David Autor. 2011. "Skills, Tasks and Technologies: Implications for Employment and Earnings." In *Handbook of Labor Economics*, vol. 4. Elsevier. https://doi.org/10.1016/S0169-7218(11)02410-5.

Acemoglu, Daron, and Pascual Restrepo. 2019. "Automation and New Tasks: How Technology Displaces and Reinstates Labor." *Journal of Economic Perspectives* 33 (2): 3–30. https://doi.org/10.1257/jep.33.2.3.

Acemoglu, Daron, and Pascual Restrepo. 2020. "Robots and Jobs: Evidence from US Labor Markets." *The Journal of Political Economy* 128 (6): 1–57. https://doi.org/10.1086/705716.

Agrawal, Ajay K., John McHale, and Alexander Oettl. 2026. *Enhancing Worker Productivity Without Automating Tasks: A Different Approach to AI and the Task-Based Model*. Working Paper No. 34781. National Bureau of Economic Research. https://doi.org/10.3386/w34781.

American Bar Association. 2025. *The Legal Industry Report 2025*. Law Technology Today. May 6. https://www.americanbar.org/groups/law_practice/resources/law-technology-today/2025/the-legal-industry-report-2025/.

Angrist, Joshua D., Guido W. Imbens, and Donald B. Rubin. 1996. "Identification of Causal Effects Using Instrumental Variables." *Journal of the American Statistical Association* 91 (434): 444–55. https://doi.org/10.1080/01621459.1996.10476902.

Autor, David H. 2024. *Applying AI to Rebuild Middle Class Jobs*. Working Paper No. 32140. National Bureau of Economic Research. https://doi.org/10.3386/w32140.

Autor, David H., Frank Levy, and Richard J. Murnane. 2003. "The Skill Content of Recent Technological Change: An Empirical Exploration." *The Quarterly Journal of Economics* 118 (4): 1279–333. https://doi.org/10.1162/003355303322552801.

Bauer, Emmanuel, Dominik Stammbach, Nianlong Gu, and Elliott Ash. 2023. "Legal Extractive Summarization of U.S. Court Opinions." *CEUR Workshop Proceedings* 3594. https://ceur-ws.org/Vol-3594/paper1.pdf.

Bresnahan, F., and M. Trajtenberg. 1995. "General Purpose Technologies 'Engines of Growth'?" *Journal of Econometrics* 65 (1): 83–108. https://doi.org/10.1016/0304-4076(94)01598-T.

Brynjolfsson, Erik, Danielle Li, and Lindsey R. Raymond. 2023. *Generative AI at Work*. Working Paper No. 31161. National Bureau of Economic Research. https://doi.org/10.3386/w31161.

Charlotin, Damien. n.d. "AI Hallucination Cases." Accessed January 28, 2026. https://www.damiencharlotin.com/hallucinations/.

Chien, Colleen V., and Miriam Kim. 2025. "Generative AI and Legal Aid: Results from a Field Study and 100 Use Cases to Bridge the Access to Justice Gap." *Loyola of Los Angeles Law Review* 57 (4): 903–88. https://digitalcommons.lmu.edu/llr/vol57/iss4/2.

Choi, Jonathan H., Amy Monahan, and Daniel B. Schwarcz. 2024. "Lawyering in the Age of Artificial Intelligence." *Minnesota Law Review* 109 (1): 147–218. https://doi.org/10.24926/265535.4225.

Choi, Jonathan H., and Daniel B. Schwarcz. 2025. "AI Assistance in Legal Analysis: An Empirical Study." *Journal of Legal Education* 73 (2): 384–420. https://jle.aals.org/home/vol73/iss2/5/.

Conley, Anna. 2025. "Understanding the Duty of Competence for Attorneys Using Generative Ai." *North Carolina Journal of Law & Technology* 27: 1–26. https://doi.org/10.2139/ssrn.5053423.

Cui, Kevin Zheyuan, Mert Demirer, Sonia Jaffe, Leon Musolff, Sida Peng, and Tobias Salz. 2025. "The Effects of Generative AI on High-Skilled Work: Evidence from Three Field Experiments with Software Developers." Preprint, Massachusetts Institute of Technology, Department of Economics, February. https://economics.mit.edu/sites/default/files/2024-01/draft_copilot_experiments.pdf.

Dahl, Matthew, Varun Magesh, Mirac Suzgun, and Daniel E. Ho. 2024. "Large Legal Fictions: Profiling Legal Hallucinations in Large Language Models." *Journal of Legal Analysis* 16 (1): 64–93. https://doi.org/10.1093/jla/laae003.

Dell'Acqua, Fabrizio, Edward McFowland, Ethan R. Mollick, et al. 2023. "Navigating the Jagged Technological Frontier: Field Experimental Evidence of the Effects of AI on Knowledge Worker Productivity and Quality." *SSRN Electronic Journal*, ahead of print. https://doi.org/10.2139/ssrn.4573321.

Denniston, Alex, and Peter Duffy. 2026. "We Trained 3,000 Lawyers in Generative AI. Here's What We Learned." *Bloomberg Law*, March 6. https://news.bloomberglaw.com/legal-exchange-insights-and-commentary/we-trained-3-000-lawyers-in-generative-ai-heres-what-we-learned.

Eloundou, Tyna, Sam Manning, Pamela Mishkin, and Daniel Rock. 2023. "GPTs Are GPTs: An Early Look at the Labor Market Impact Potential of Large Language Models." arXiv:2303.10130. Preprint, arXiv, August 21. https://doi.org/10.48550/arXiv.2303.10130.

Engel, Christoph, and Johannes Kruse. 2025. *LLM as a Law Professor: Having a Large Language Model Write a Commentary on Freedom of Assembly*. MPI Collective Goods Discussion Paper. https://papers.ssrn.com/sol3/papers.cfm?abstract_id=5435276.

Frangakis, Constantine E., and Donald B. Rubin. 2002. "Principal Stratification in Causal Inference." *Biometrics* 58 (1): 21–29. https://doi.org/10.1111/j.0006-341X.2002.00021.x.

Freitas, Julian. 2025. "Why People Resist Embracing AI." *Harvard Business Review*, January. https://hbr.org/2025/01/why-people-resist-embracing-ai.

Humlum, Anders, and Emilie Vestergaard. 2024. *The Adoption of ChatGPT*. IZA Discussion Paper No. 16992. IZA Institute of Labor Economics.

Imbens, Guido W., and Charles F. Manski. 2004. "Confidence Intervals for Partially Identified Parameters." *Econometrica* 72 (6): 1845–57. https://doi.org/10.1111/j.1468-0262.2004.00555.x.

Lam, Kwok-Yan, Victor C. W. Cheng, and Zee Kin Yeong. 2023. "Applying Large Language Models for Enhancing Contract Drafting." *Proceedings of the Third International Workshop on Artificial Intelligence and Intelligent Assistance for Legal Professionals in the Digital Workspace* 3423 (June). https://ceur-ws.org/Vol-3423/paper7.pdf.

LexisNexis. 2024. *The State of GenAI in Law: A LexisNexis Report*. https://www.lexisnexis.com/pdf/genai-report.pdf.

Nielsen, Aileen, Stavroula Skylaki, Milda Norkute, and Alexander Stremitzer. 2024. "Building a Better Lawyer: Experimental Evidence That Artificial Intelligence

Can Increase Legal Work Efficiency." *Journal of Empirical Legal Studies* 21 (4): 979–1022. https://doi.org/10.1111/jels.12396.

Noy, Shakked, and Whitney Zhang. 2023. "Experimental Evidence on the Productivity Effects of Generative Artificial Intelligence." *Science* 381 (6654): 187–92. https://doi.org/10.1126/science.adh2586.

Peng, Sida, Eirini Kalliamvakou, Peter Cihon, and Mert Demirer. 2023. "The Impact of AI on Developer Productivity: Evidence from GitHub Copilot." arXiv:2302.06590. Preprint, arXiv, February 13. https://doi.org/10.48550/arXiv.2302.06590.

Rubin, Donald B. 1974. "Estimating Causal Effects of Treatments in Randomized and Nonrandomized Studies." *Journal of Educational Psychology* 66 (5): 688–701. https://doi.org/10.1037/h0037350.

Rubin, Donald B. 1976. "Inference and Missing Data." *Biometrika* 63 (3): 581–92. https://doi.org/10.1093/biomet/63.3.581.

Rubin, Donald B. 1997. "Estimating Causal Effects from Large Data Sets Using Propensity Scores." *Annals of Internal Medicine* 197 (8_Part_2): 757–63. https://doi.org/10.7326/0003-4819-127-8_Part_2-199710151-00064.

Savelka, Jaromir. 2023. "Unlocking Practical Applications in Legal Domain: Evaluation of GPT for Zero-Shot Semantic Annotation of Legal Texts." *Proceedings of the*

*Nineteenth International Conference on Artificial Intelligence and Law*, June 19, 447–51. https://doi.org/10.1145/3594536.3595161.

Schwarcz, Daniel, and Jonathan H. Choi. 2023. “AI Tools for Lawyers: A Practical Guide.” *Minnesota Law Review Headnotes* 108 (1): 1–39. https://minnesotalawreview.org/wp-content/uploads/2023/10/FL1-Choi-Schwarcz.pdf.

Schwarcz, Daniel, Sam Manning, Patrick James Barry, David R. Cleveland, J. J. Prescott, and Beverly Rich. 2025. “AI-Powered Lawyering: AI Reasoning Models, Retrieval Augmented Generation, and the Future of Legal Practice.” Preprint, SSRN. https://doi.org/10.2139/ssrn.5162111.

Splawa-Neyman, Jerzy, D. M. Dabrowska, and T. P. Speed. 1990. “On the Application of Probability Theory to Agricultural Experiments. Essay on Principles. Section 9.” *Statistical Science* 5 (4). https://doi.org/10.1214/ss/1177012031.

Thomson Reuters. 2025. *How AI Is Transforming the Legal Profession*. https://legal.thomsonreuters.com/blog/how-ai-is-transforming-the-legal-profession/.